\documentclass[aps,prl,preprint,nopacs,superscriptaddress]{revtex4}
\usepackage{setspace}
\usepackage{lipsum}
\usepackage{graphicx}
\usepackage{verbatim}
\usepackage{relsize}
\usepackage{mathrsfs}
\usepackage{color}
\usepackage{amsmath,amsfonts,amssymb}
\usepackage{graphicx}

\def \beq {\begin{equation}}
\def \eeq {\end{equation}}
\begin{document}
\title{Electrically switchable Berry curvature dipole in the monolayer topological insulator WTe$_2$}

\author{Su-Yang Xu$^*$}\affiliation {Department of Physics, Massachusetts Institute of Technology, Cambridge, Massachusetts 02139, USA}
\author{Qiong Ma\footnote{These authors contributed equally to this work.}}\affiliation{Department of Physics, Massachusetts Institute of Technology, Cambridge, Massachusetts 02139, USA}
\author{Huitao Shen}\affiliation {Department of Physics, Massachusetts Institute of Technology, Cambridge, Massachusetts 02139, USA}
\author{Valla Fatemi}\affiliation{Department of Physics, Massachusetts Institute of Technology, Cambridge, Massachusetts 02139, USA}
\author{Sanfeng Wu}\affiliation{Department of Physics, Massachusetts Institute of Technology, Cambridge, Massachusetts 02139, USA}
\author{Tay-Rong Chang}
\affiliation{Department of Physics, National Cheng Kung University, Tainan 701, Taiwan}
\author{Guoqing Chang}
\affiliation{Institute of Physics, Academia Sinica, Taipei 11529, Taiwan}
\author{Andr\'es M. Mier Valdivia}\affiliation{Department of Physics, Massachusetts Institute of Technology, Cambridge, Massachusetts 02139, USA}
\author{Ching-Kit Chan}\affiliation{Department of Physics and Astronomy, University of California Los Angeles, Los Angeles, California 90095, USA}
\author{Quinn D. Gibson}
\affiliation{Department of Chemistry, Princeton University, Princeton, New Jersey 08544, USA}
\author{Kenji Watanabe}
\affiliation{National Institute for Materials Science, Namiki 1 -1, Tsukuba, Ibaraki 305 -0044, Japan}
\author{Takashi  Taniguchi}
\affiliation{National Institute for Materials Science, Namiki 1 -1, Tsukuba, Ibaraki 305 -0044, Japan}
\author{Hsin Lin}
\affiliation{Institute of Physics, Academia Sinica, Taipei 11529, Taiwan}
\author{Robert J. Cava}
\affiliation{Department of Chemistry, Princeton University, Princeton, New Jersey 08544, USA}
\author{Liang Fu}\affiliation {Department of Physics, Massachusetts Institute of Technology, Cambridge, Massachusetts 02139, USA}
\author{Nuh Gedik$^{\dag}$}\affiliation {Department of Physics, Massachusetts Institute of Technology, Cambridge, Massachusetts 02139, USA}
\author{Pablo Jarillo-Herrero\footnote{Corresponding authors (emails): gedik@mit.edu and pjarillo@mit.edu }}\affiliation {Department of Physics, Massachusetts Institute of Technology, Cambridge, Massachusetts 02139, USA}

\date{\today}

\begin{abstract}
Recent experimental evidence for the quantum spin Hall (QSH) state in monolayer WTe$_2$ has bridged two of the most active fields of condensed matter physics, 2D materials and topological physics \cite{Qian2014, Zheng2016, Fei2017, Tang2017, jia2017direct, Sanfeng_Talk}. This 2D topological crystal also displays unconventional spin-torque \cite{macneill2016} and gate-tunable superconductivity \cite{Sanfeng_Talk}. While the realization of QSH \cite{Qian2014, Zheng2016, Fei2017, Tang2017, jia2017direct, Sanfeng_Talk} has demonstrated the nontrivial topology of the electron wavefunctions of monolayer WTe$_2$, the geometrical properties of the wavefunction, such as the Berry curvature, remain unstudied. On the other hand, it has been increasingly recognized that the Berry curvature \cite{nagaosa2010anomalous, xiao2010berry} plays an important role in multiple areas of condensed matter physics including nonreciprocal electron transport \cite{moore2010confinement, sodemann2015quantum, rectification, Edelstein}, enantioselective optical responses \cite{magnetochiral, Xiaodongreview, mak2016photonics,  de2017quantized}, chiral polaritons \cite {basov2016polaritons, low2017polaritons} and even unconventional superconductivity \cite{gradhand2014berry}. Here we utilize mid-infrared optoelectronic microscopy to investigate the Berry curvature in monolayer WTe$_2$. By optically exciting electrons across the inverted QSH gap, we observe an in-plane circular photogalvanic current even under normal incidence. The application of an out-of-plane displacement field further systematically controls the direction and magnitude of the photocurrent. Our observed photocurrent reveals a novel Berry curvature dipole that arises from the nontrivial wavefunctions near the inverted gap edge. These previously unrealized Berry curvature dipole and strong electric field effect are uniquely enabled by the inverted band structure and tilted crystal lattice of monolayer WTe$_2$. Such an electrically switchable Berry curvature dipole opens the door to the observation of a wide range of quantum geometrical phenomena,  such as quantum nonlinear Hall \cite{sodemann2015quantum}, orbital-Edelstein \cite{Edelstein} and chiral polaritonic effects \cite{basov2016polaritons, low2017polaritons}.  
\vspace{0.6cm}

\end{abstract}
\pacs{}

\maketitle

One of the early landmarks of condensed matter physics was the classification of metals, insulators and semiconductors by studying the energy-momentum dispersion (band structure) of the electrons in crystalline solids. Despite such remarkable success, the quantum nature means that the electron states can only be fully described by their quantum wavefunctions, whereas the band structure only concerns the energy and momentum eigenvalues of the wavefunctions. Therefore, a central question in modern condensed matter physics is whether there exist new phenomena that arise from other properties of quantum wavefunctions beyond the band structure \cite{nagaosa2010anomalous, xiao2010berry}. For instance, the study of the global (topological) properties of electronic wavefunctions continues to give rise to novel topological phases including the QSH states, 3D topological insulators and Weyl semimetals. These topological materials feature robust surface/edge states and often exhibit protected transport and optical properties. Another direction is to study the local (geometrical) properties of wavefunctions. One important property is the local curvature of the wavefunction, defined as the Berry curvature (BC) \cite{nagaosa2010anomalous}. Originally employed to explain the anomalous Hall conductivities of ferromagnets \cite{nagaosa2010anomalous}, the importance of BC is increasingly recognized in a wide range of areas in condensed matter physics, including nonlocal transport and chiral optical responses in noncentrosymmetric metals and semiconductors \cite{Xiaodongreview, mak2016photonics, son2013chiral, burkov2015negative, moore2010confinement, sodemann2015quantum, rectification, Edelstein, magnetochiral, de2017quantized}, skyrmion transport in noncentrosymmetric magnets \cite{nagaosa2013topological}, unconventional pairing in superconductors \cite{gradhand2014berry}, and topological plasmonic and excitonic polaritons \cite{basov2016polaritons, low2017polaritons}. Moreover, the interplay between topology and Berry curvature, although of great fundamental interest \cite{de2017quantized, morimoto2016chiral, Edelstein}, has been rarely explored. This is because most topological materials have zero Berry curvature in their bulk electronic states due to inversion symmetry, whereas materials with nonzero Berry curvature (e.g., monolalyer MoS$_2$ or gapped graphene) are mostly topologically trivial. In general, the research on understanding the effects of quantum geometry in novel materials is developing rapidly both in theory and in experiments \cite{moore2010confinement, nagaosa2010anomalous, nagaosa2013topological, gradhand2014berry, basov2016polaritons, low2017polaritons, Xiaodongreview, mak2016photonics, son2013chiral, burkov2015negative, sodemann2015quantum, rectification, Edelstein, de2017quantized, magnetochiral}. It is of importance to find new materials with novel quantum geometrical and topological properties.

Bulk WTe$_2$ crystals were found to show a large, non-saturating magnetoresistance \cite{ali2014large} and were proposed as type-II Weyl semimetals \cite{soluyanov2015type}. More recently, monolayer WTe$_2$ was experimentally identified as a QSH insulator \cite {Fei2017, Tang2017, jia2017direct, Sanfeng_Talk} following the prediction by Qian et al.\cite{Qian2014}. Spin-torque and gate-tunable superconductivity were also observed in monolayer WTe$_2$ \cite{macneill2016, Sanfeng_Talk}. While the realization of QSH in monolayer WTe$_2$ demonstrates the nontrivial topological invariant of its quantum wavefunction \cite {Fei2017, Tang2017, jia2017direct, Sanfeng_Talk}, the geometrical properties of the wavefunction remain entirely unstudied theoretically and experimentally. Meanwhile, although monolayer WTe$_2$ has been studied by electronic transport \cite{Fei2017, Sanfeng_Talk}, angle-resolved photoemission (ARPES) \cite{Tang2017}, and scanning tunneling microscopy (STM) \cite{Tang2017, jia2017direct}, optical studies are lacking. 

The circular photogalvanic effect (CPGE) is the generation of electrical currents via circular polarized (CP) light. Previous CPGE experiments on semiconductor (e.g. GaAs) heterostructures \cite{ganichev2003spin, wittmann2010circular}, topological insulators \cite{mciver2012control} and semiconducting transition metal dichalcoginides \cite{yuan2014generation} have attracted great interest because of the ability to optically generate spin polarized electrical currents. On the other hand, a systematic microscopic mechanism of the these CPGEs is usually difficult to achieve because the complex optical processes often involve many bands. However, recent theoretical advances \cite{de2017quantized} have shown that, when the inter-band transition only involves the lowest two bands, the CPGE in 3D bulk materials has a concise microscopic origin that arises from the nontrivial BC. As a result, aside from being a novel and potentially useful phenomenon, the CPGE under such conditions becomes a powerful probe of important wavefunction properties of a bulk material, such as the chirality and the topological charge of the Weyl nodes \cite{de2017quantized, ma2017direct}. Here, we show that the CPGE under the same conditions in 2D also has a clear BC origin. 

\begin{figure*}[t]
\includegraphics[width=16cm]{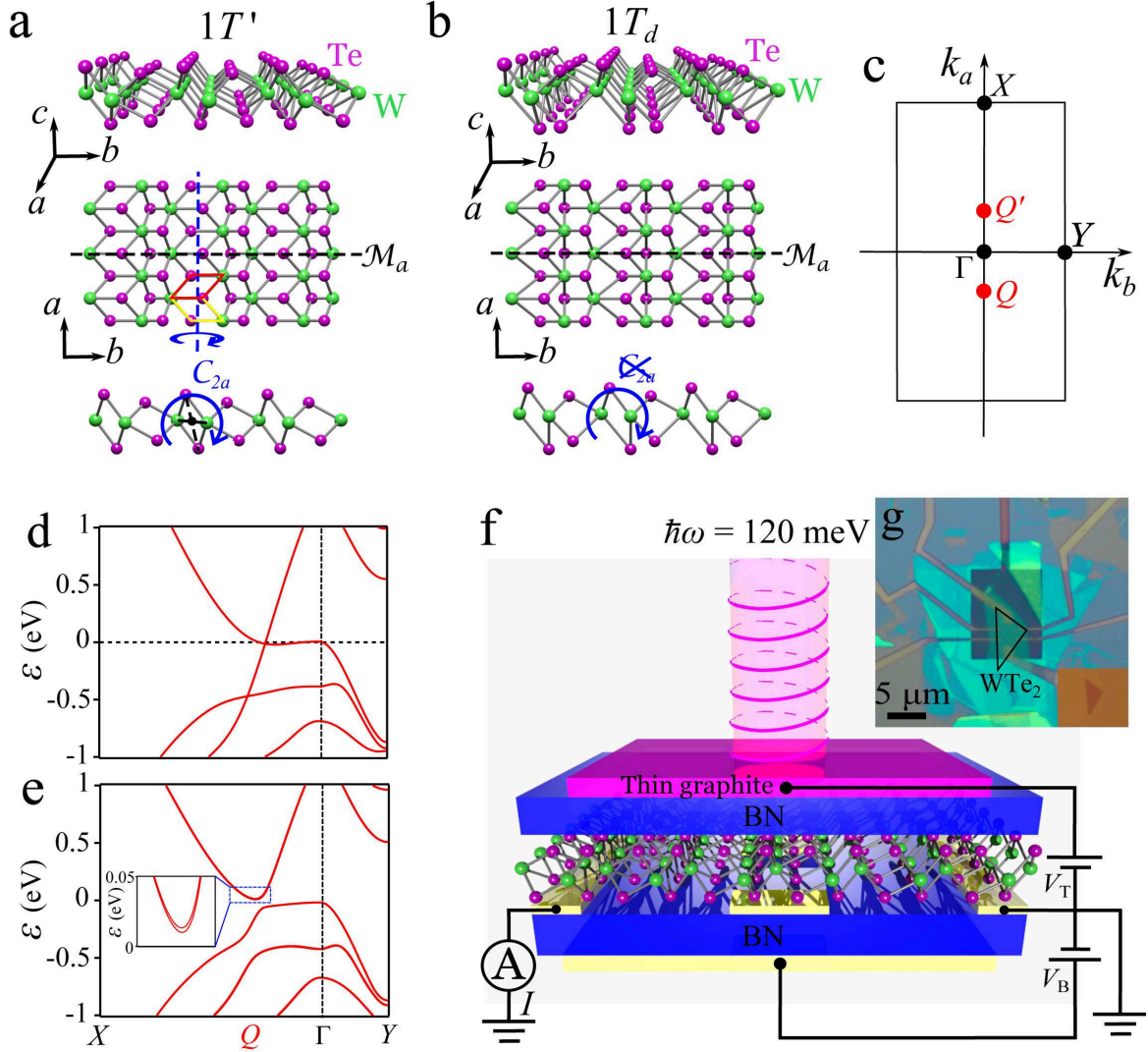}
\caption{{\bf Crystal and electronic structures of monolayer WTe$_2$.} \textbf{a,} The $1T'$ structure of monolayer WTe$_2$ consists of a mirror plane $\mathcal{M}_a$ and a screw rotate symmetry $\mathcal{C}_{2a}$. $\mathcal{C}_{2a}$ is nonsymmorphic: it involves a $180^{\circ}$ rotation about $\hat{a}$ and a $\frac{a}{2}$ translation along $\hat{a}$, as depicted by the yellow and red parallelograms. \textbf{b,} The $1T_d$ structure only has the mirror plane $\mathcal{M}_a$. The rotate symmetry $\mathcal{C}_{2a}$ is broken (exaggerated). \textbf{c,} The first Brillouin zone with important momenta noted. \textbf{d,e,} band structure of monolayer WTe$_2$ without and with spin-orbit coupling. \textbf{f,} Schematic experimental setup for detecting the mid-infrared circular photogalvanic effect (CPGE) on a dual-gated monolayer WTe$_2$ device. \textbf{g,} Optical image of a dual-gated monolayer WTe$_2$ device. Scale bar: $5 \mu$m. }
\label{Fig1}
\end{figure*}

The monolayer WTe$_2$ lattice can be described by two possible structural phases, $1T'$ and $1T_d$. The inversion-symmetric $1T'$ structure has been widely assumed by previous works \cite{Qian2014, Zheng2016, Fei2017, Tang2017}. This structure has two independent symmetries: the mirror symmetry $\mathcal{M}_a$ and the two-fold screw rotational symmetry $C_{2a}$ (Fig.~\ref{Fig1}a), whose combination gives rise to the inversion symmetry of the $1T'$ phase. The $1T_d$ structure (Fig.~\ref{Fig1}b), which is defined here as the monolayer directly isolated from the inversion-breaking bulk $T_d$ WTe$_2$ lattice structure (see details in SI. V.1), deviates slightly from $1T'$. In $1T_d$, $\mathcal{M}_a$ is preserved but $C_{2a}$ is weakly broken. As a result, $1T_d$ actually breaks inversion symmetry, which affects its electronic structure. The low-energy band structure of monolayer WTe$_2$ without spin-orbit coupling (SOC) features tilted 2D Dirac fermions at the $Q$ and $Q'$ points (Figs.~\ref{Fig1}c,d) \cite{muechler2016topological}. The inclusion of SOC leads to an inverted, indirect QSH gap, where the valence band top and conduction band bottom are located at $\Gamma$ and $Q$($Q'$), respectively. The weak inversion breaking of $1T_d$ further induces a small spin splitting near the bottom of the conduction band (Fig.~\ref{Fig1}e inset). 

Here we use a mid-infrared scanning photocurrent microscope equipped with a CO$_2$ laser ($\lambda=10.6$ $\mu{m}$, $\hbar\omega\simeq120$ meV) to detect the CPGE induced by the inter-band transition across the inverted QSH gap near $Q$ and $Q'$. We have fabricated high-quality, encapsulated, dual-gated monolayer WTe$_2$ devices in the Hall bar geometry (Figs.\ref{Fig1}f,g). The dual gates allow us to independently vary the charge density $n$ and the displacement field $\vec{D}$ (see details in the methods section).

\begin{figure*}[t]
\includegraphics[width=14cm]{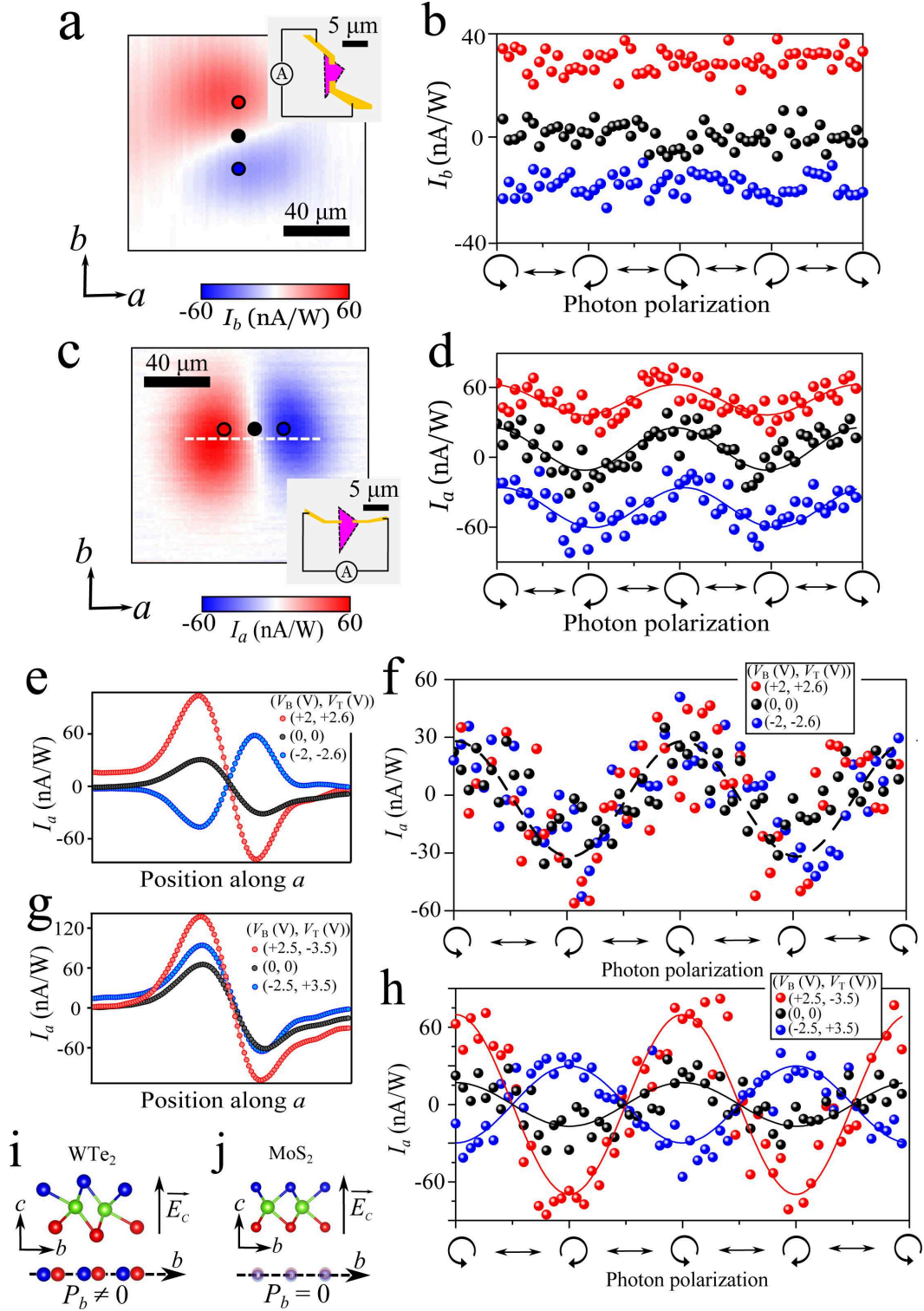}
\caption{
{\bf Observation of circular photogalvanic effect (CPGE) in monolayer WTe$_2$.} \textbf{a,} Photocurrent along $\hat{b}$ ($I_{\hat{b}}$) with a fixed polarization (RCP) while the light spot is shifted in $\hat{a}-\hat{b}$ plane. \textbf{b,} Polarization-dependent $I_{\hat{b}}$ with the light spot fixed at the black dot in panel (a).}
\label{Fig2}
\end{figure*}

We first present our data at $T=150$ K without external displacement fields. Figures~\ref{Fig2}a,c show the measured photocurrents along two orthogonal directions ($I_{\hat{a}}$ and $I_{\hat{b}}$) as a function of the laser position. The photocurrent changes sign as one moves the light spot from one contact to the opposite. This spatial pattern reveals the photo-thermal current \cite{mciver2012control, yuan2014generation} along both $\hat{a}$ and $\hat{b}$, which is due to the different Seebeck coefficients of WTe$_2$ and metal contacts. By contrast, the polarization dependence of the photocurrent along the two directions (Figs.~\ref{Fig2}b,d) is distinctly different. $I_{\hat{a}}$ (Fig.~\ref{Fig2}d) shows a significant modulation with light polarization, which reaches maximum for right CP light, minimum for left CP light, and zero for linearly polarized light. This pattern clearly demonstrates the existence of CPGE along $\hat{a}$. $I_{\hat{b}}$, on the other hand, shows no observable dependence on polarization (Fig.~\ref{Fig2}b). To further distinguish the CPGE from the photo-thermal effect, we study the dependence with charge density, without applying displacement fields. As seen in Figs.~\ref{Fig2}e,f, the CPGE remains unchanged within a relatively large charge density range ($|n|\leq10^{13}$ cm$^{-2}$), whereas the photo-thermal current changes sign as one varies the doping from electron-like to hole-like. Since any CPGE would vanish with inversion symmetry, the observed CPGE without displacement field suggests the inversion breaking $1T_d$ phase as the actual structure of monolayer WTe$_2$. On the other hand, this weak inversion breaking could also come from the dielectric environment, although this is less likely because of the (almost) symmetric hBN encapsulation and the supposedly weak interaction between WTe$_2$ and hBN due to their very different lattices. In SI.I, we further show that the directional dependence of the CPGE is consistent with the symmetry analysis. 

We now study the dependence of the CPGE with an external, out-of-plane, displacement field $\vec{D}$. As shown by the red curve in Fig.~\ref{Fig2}h, the application of a displacement field significantly increases the CPGE. By contrast, as we reverse the direction of the displacement field, the CPGE (blue curve) flips sign. Such electrical switching and sign-reversal have not been achieved homogeneously throughout a 3D bulk system \cite{ganichev2003spin, wittmann2010circular, mciver2012control, yuan2014generation, ma2017direct, dhara2015voltage}, as the electrical gating only affects the surface or interface region of a bulk crystal. The strong dependence of the in-plane CPGE with the out-of-plane $\vec{D}$ reveals a previously uncharacterized field effect in monolayer WTe$_2$. Conventionally, e.g. in graphene and MoS$_2$, an out of plane $\vec{D}$ only causes an out-of-plane (+$\hat{c}$ to $-\hat{c}$) polarity \cite{wu2013electrical, taychatanapat2010electronic, Xiaodongreview, mak2016photonics} (Fig.~\ref{Fig2}j). By contrast, because of the distinct crystal structure of monolayer WTe$_2$ that features a tilted parallelogram on the $\hat{b}-\hat{c}$ plane (Fig.~\ref{Fig2}i), an out of plane $\vec{D}$ can give rise to an in-plane polarity along $\hat{b}$, which eventually modulates the in-plane CPGE. In SI. I, we further show the symmetry analysis of the observed CPGE with the displacement field. 

We then study the CPGE at low temperatures ($T=20$ K). Guided by the gate map of the four-probe electrical resistance $R_{xx}(V_T,V_B)$ (Fig.~\ref{Fig3}a), we are able to tune the displacement field over a wide range  while keeping charge density invariant. As shown in Fig.~\ref{Fig3}b, both the direction and the magnitude of the CPGE can be controlled and modulated as a function of the displacement field. Moreover, in contrast to the measurements at $T=150$ K in Fig.~\ref{Fig2}, we find no observable CPGE within a finite range of small displacement fields ($\arrowvert\vec{D}\arrowvert\leq$ $0.5$ V$\cdot$nm$^{-1}$) at $T=20$ K (Figs.~\ref{Fig3}b-d). To further understand the contrasting behaviors at $T=150$ K and $T=20$ K, we study the temperature dependence. The CPGE at large $\vec{D}$ is strong at both low and high temperatures (Figs.~\ref{Fig2}h,~\ref{Fig3}b-d). By contrast, the CPGE at $\vec{D}=0$ is weak at low temperatures and increases significantly at $T\sim100$ K (Fig.~\ref{Fig3}e). These observations collectively indicate a monolayer WTe$_2$ band gap that is dependent on both temperature and the displacement field. Specifically, as we lower the temperature from 150 K to 20 K, the direct band gap (the energy gap between the lowest conduction and the highest valence bands at a fixed $\vec{k}$) increases and becomes greater than our photon energy $\hbar\omega=120$ meV, which hinders the inter-band transition process. The displacement field dependence of the band gap can be directly reproduced by our first-principles calculated band structures of monolayer WTe$_2$ (Figures.~\ref{Fig3}f-i). Moreover, our calculations show that the displacement field induces a strong Berry curvature concentrated near the inverted gap edge. 

\color{black} 

We now turn to the microscopic mechanism of the observed CPGE. We consider the scenario where the optical inter-band transition only involves the lowest conduction and highest valence bands. Under such conditions, we show below that the optical selection rules and the CPGE directly depend on the BC $\Omega(\vec{k})$. In Figs.~\ref{Fig4}a,c and ~\ref{Fig3}h, we find the following important characteristics for the BC of monolayer WTe$_2$ at a fixed field ($E^{\textrm{THY}}=+0.5$ V/nm): (1) $\Omega_c(\vec{k})$ exhibits clear hotspots near the $Q$ and $Q'$ points where the direct energy gap is minimum. (2) At all energies, the BC shows a bipolar configuration about the mirror plane $\mathcal{M}_a$. Inspired by the concept considered in \cite{sodemann2015quantum, zhang2017berry} for intra-band physics, we define an inter-band BC dipole ($\vec{\mathlarger{\mathlarger{\Lambda}}}^{\Omega}$). 

\begin{equation}
\vec{\mathlarger{\mathlarger{\Lambda}}}^{\Omega}=\oint d\vec{k} \times \vec{\Omega}(\vec{k}),
\label{Dipole}
\end{equation} where the closed loop integral ($\oint d\vec{k}$) is defined along the $k$-contours that correspond to an $\hbar \omega$ transition between bands 2 and 3. Therefore, $\vec{\mathlarger{\mathlarger{\Lambda}}}^{\Omega}$ measures the degree of polarity of the BC texture on these $k$-contours (Fig.~\ref{Fig4}a) corresponding to an $\hbar \omega$ inter-band transition. Combining Eq.~\ref{Dipole} with the fact that $\Omega_c(k_a,k_b)=-\Omega_c(-k_a,k_b)$ enforced by the mirror plane $\mathcal{M}_a$, it is evident that one has $\mathlarger{\mathlarger{\Lambda}}^{\Omega}_a\neq0, \mathlarger{\mathlarger{\Lambda}}^{\Omega}_b=0$ for monolayer WTe$_2$. We note that the following two factors are important to generate a nonzero BC dipole $\mathlarger{\mathlarger{\Lambda}}$ in monolayer WTe$_2$: First, around each contour, the BC magnitude is not uniform; Second, there are only two contours with opposite BC (Fig.~\ref{Fig4}a).

\color{black} 
We show how such a polar BC texture leads to our observed CPGE. The generation of CPGE with normal incident light consists of two steps: (1) CP light excites inter-band transitions with an optical selection rule. (2) The optically excited electrons and holes travel at their respective group velocities, leading to a nonzero CPGE current. In the case where only two bands participate in the inter-band process, it has been theoretically shown \cite{xiao2007valley, de2017quantized, wu2013electrical} that the difference between the inter-band transition rate for RCP and LCP light, $\mathcal{V}(\vec{k})=\arrowvert{\mathcal{P}^{\textrm{RCP}}(\vec{k})}\arrowvert^2-\arrowvert{\mathcal{P}^{\textrm{LCP}}(\vec{k})}\arrowvert^2$, is directly proportional to the BC, i.e., $\mathcal{V}(\vec{k})\propto\Omega(\vec{k})$ (see SI.III for derivations). Therefore, the inter-band transition near $Q$ and $Q'$ in monolayer WTe$_2$ selects opposite CP light because of their opposite BCs. Under this condition, the CPGE with normal incident light can be expressed concisely in terms of the BC dipole (see SI.III for derivations)

\begin{align}
\vec{J}^{\textrm{CPGE}}&=\frac{e^3\tau}{\pi\hbar^2}\,\mathcal{I}m\bigg[\vec{E}(-\omega)\times\hat{c}\,\,(\vec{\mathlarger{\mathlarger{\Lambda}}}^{\Omega}\cdot\vec{E}(\omega))\bigg],
\label{CPGE_Dipole}
\end{align} where $\tau$ is the relaxation time, $\vec{E}(\omega)=\frac{E_0}{\sqrt{2}}(e^{i\omega{t}},e^{i(\omega{t}\pm\frac{\pi}{2})},0)$ describes normal incident RCP and LCP light. We see that, while the circular dichroic optical transition is possible as long as there is finite BC ($\mathcal{V}(\vec{k})\propto\Omega(\vec{k})$) \cite{Xiaodongreview, mak2016photonics, wu2013electrical}, the CPGE here can only occur in the presence of a Berry curvature dipole ($\vec{\mathlarger{\mathlarger{\Lambda}}}^{\Omega}\neq0$). The nonzero $\mathlarger{\mathlarger{\Lambda}}^{\Omega}_a$ in monolayer WTe$_2$ directly leads to the observed CPGE current along $\hat{a}$ under normal incidence. In Fig.~\ref{Fig4}f, we further study the $\vec{D}$ dependence of the BC dipole $\mathlarger{\mathlarger{\Lambda}}^{\Omega}_a$. For $|E|<0.2$ V/nm, the direct band gap is too large for an $\hbar\omega=120$ meV transition; For $|E|\geq0.2$ V/nm, $\arrowvert\vec{\mathlarger{\mathlarger{\Lambda}}}^{\Omega}\arrowvert$ increases monotonically with $|E|$. Importantly, both the trend and the order of magnitude of the calculated BC dipole (Fig.~\ref{Fig4}f) are consistent with the our experimental data (Fig.~\ref{Fig3}d, see SI. IV for details). 

We now study the origin of the observed field effect in monolayer WTe$_2$. In contrast to the very weak displacement field effect found in monolayer MoS$_2$  \cite{Xiaodongreview, mak2016photonics, wu2013electrical}, the field effect induced band splitting and BC in monolayer WTe$_2$ are selectively strong near the inverted band gap ($Q(Q')$ point) but weak elsewhere in $k$ space (Fig.~\ref{Fig3}f-i). To understand this, we calculate the real-space distribution of the wavefunction amplitude at the inverted gap edge ($Q$ point) (Fig.~\ref{Fig4}i). Interestingly, in the presence of band inversion, the wavefunction spans across the three atomic layers (Fig.~\ref{Fig4}j). As we remove the band inversion (Fig.~\ref{Fig4}g), the wavefunction becomes strongly localized near the central W atomic layer (Fig.~\ref{Fig4}h). Moreover, the field effect induced band splitting and BC are very small without the band inversion. Therefore, the uncovered unique field effect and the emergent BC hotspot near the inverted gap edge in monolayer WTe$_2$ are fingerprints of the topological band inversion. Such a topological field effect can be generalized into other QSH systems (and more broadly 2D materials with band inversions) to induce strong nontrivial BCs.

To understand the role of spin in the CP selection rule, we further show that the spin polarizations of bands 2 and 3 (also 1 and 4) are along the same direction (Fig.~\ref{Fig4}e, see SI.II for more details). The completely overlapping spin wavefunctions between bands 2 and 3 (Fig.~\ref{Fig4}e) demonstrate (from a different perspective) that the CP light selection rule between these two bands arises purely from the BC. Moreover, within a constant energy contour, the spin texture shows a canted (between $\hat{b}$ and $\hat{c}$) Zeeman-like configuration on each constant energy contour (Fig.~\ref{Fig4}d), which is different from the out-of-plane Zeeman-like spin texture in MoS$_2$. 

We highlight the key observations of monolayer WTe$_2$ in comparison to graphene and MoS$_2$: (1) The BC in monolayer WTe$_2$ is polar. By contrast, the BCs in MoS$_2$ and gapped graphene have a zero dipole, which can be seen also from the following two factors: First, around each contour at $K(K')$, the BC is three-fold symmetric; Second, there are three degenerate $K-K'$ pairs (Fig.~\ref{Fig4}b). In addition, due to the inverted band structure, the BC in monolayer WTe$_2$ forms an intense hotspot, whose amplitude is more than one order of magnitude larger than that in MoS$_2$. (2) The displacement field effect in monolayer WTe$_2$ is the first strong field effect in a monolayer crystal. This new field effect is a direct consequence of the topological band inversion and further shows an ``out-of-plane to in-plane'' coupling. By contrast, previous field effects in MoS$_2$ and graphene are only strong in thicker layers \cite{Xiaodongreview, mak2016photonics, wu2013electrical}, which mainly come from the coupling between different layers separated by the van der Waals gap.

Our results represent the first experimental demonstration of a BC dipole, which can be further controlled by electrical means.  Such a tunable BC dipole not only leads to the electrically switchable CPGE observed here, but further enables a wide range of other quantum geometrical phenomena, such as magnetochiral \cite{magnetochiral}, quantum nonlinear Hall \cite{sodemann2015quantum}, rectification \cite{rectification}, and orbital-Edelstein \cite{Edelstein} effects. For all these phenomena, a nonzero BC is insufficient whereas a BC dipole is truly required. Many of these phenomena have not been observed in any solid state system and are therefore of great fundamental interest. Besides the above single-particle phenomena, the CP light selection rule means that monolayer WTe$_2$, analogous to MoS$_2$ and gapped graphene, can be used to realize chiral edge plasmons \cite{basov2016polaritons, low2017polaritons}. It is worth noting that monolayer WTe$_2$ has topological edge states, whose role in plasmon physics is unstudied even theoretically. Further, the effect of nontrivial BC and BC dipole in the gate-induced superconductivity \cite{Sanfeng_Talk} awaits exploration \cite{gradhand2014berry}. More broadly, the uncovered BC dipole, along with the previous observations \cite{Zheng2016, Fei2017, Tang2017, jia2017direct, macneill2016, Sanfeng_Talk}, establish monolayer WTe$_2$ as an extremely rich, atomically thin platform to explore topological physics, quantum geometrical physics, unconventional superconductivity as well as their interplays.

\vspace{1cm}

\small
\begin{singlespacing}
\textbf{Methods}

\textbf{Device fabrication:} Our fabrication of the dual-gated monolayer WTe$_2$ devices consists of two phases. Phase I was done under ambient conditions: local bottom PdAu gates were first defined on the standard Si/SiO$_2$ substrates. A suitable hexagonal hexagonal BN (hBN) flake was exfoliated onto a separate Si/SiO$_2$ substrate, picked up using a polymer-based dry transfer technique and placed onto the pre-patterned local bottom gate. Electrical contacts (PdAu, $\sim 20$ nm thick) in a Hall bar geometry were deposited onto the bottom hBN flake with $e$-beam lithography and metal deposition. Phase II was done fully inside the glovebox with argon environment. Monolayer WTe$_2$ flakes were exfoliated from a bulk crystal onto Si/SiO$_2$ chip. Thin graphite (as top gate electrode, $\sim 10$ nm), hBN ($\sim 10$ nm) and monolayer WTe$_2$ were sequentially picked up and then transferred onto the local bottom gate/hBN/contact substrate. Extended leads connecting the top gate graphene to wire bonding pads were pre-made together with the metal contacts in Phase I. In such a dual-gated device, the charge density can be obtained by $n=\frac{\epsilon_0\epsilon^{\textrm{hBN}}}{e}(V_T/h_T+V_B/h_B$). The displacement field is determined by $\vec{D}=(\epsilon^{\textrm{hBN}}V_T/h_T-\epsilon^{\textrm{hBN}}V_B/h_B)/2$ \cite{taychatanapat2010electronic}. Here $n$ is the charge density, $\vec{D}$ is the externally applied displacement field, $V_T$ ($V_B$) are the bias voltages, and $\epsilon^{\textrm{hBN}}=3$, $h_T=10$ nm, and $h_B=8$ nm are relative dielectric constant and thicknesses of the top ($T$) and bottom ($B$) hBN layers of the presented device, respectively.

\textbf{Mid-infrared scanning photocurrent microscopy:} The fabricated device was wire bonded onto a chip carrier and placed in an optical scanning microscope setup that combines electronic transport measurements with light illumination. The laser source is a temperature-stablized CO$_2$ laser ($\lambda = 10.6$ $\mu$m $\hbar\omega = 120$ meV). A focused beam spot (diameter $d \simeq 50$ $\mu$m) is scanned (using a two axis piezo-controlled scanning mirror) over the entire sample and the current is recorded at the same time to form a color map of photocurrent as a function of spatial positions. Reflected light from the sample is collected to form a simultaneous reflection image of the sample. The absolute location of the photo-induced signal is therefore found by comparing the photocurrent map to the reflection image. The light is first polarized by a polarizer and the chirality of light is further modulated by a rotatable quarter-wave plate. 

\vspace{1cm}
\textbf{First-principles calculations:} First-principles calculations were performed by the OPENMX code within the framework of the generalized gradient approximation of density functional theory \cite{DFT4}.

\textbf{Data availability:} The data that support the plots within this paper and other findings of this study are available from the corresponding author upon reasonable request.

\textbf{Acknowledgement:} We acknowledge Yuxuan Lin and Tom\'as Palacios for their assistance on measurements. NG and SYX acknowledge support from U.S. Department of Energy, BES DMSE, award number DE-FG02-08ER46521 (initial planning), the Gordon and Betty Moore FoundationÕs EPiQS Initiative through Grant GBMF4540 (data analysis),  and in part from the MRSEC Program of the National Science Foundation under award number DMR - 1419807 (data taking and manuscript writing). Work in the PJH group was partly supported by the Center for Excitonics, an Energy Frontier Research Center funded by the US Department of Energy (DOE), Office of Science, Office of Basic Energy Sciences under Award Number DESC0001088 (fabrication and measurement) and partly through AFOSR grant FA9550-16-1-0382 (data analysis), as well as the Gordon and Betty Moore Foundation's EPiQS Initiative through Grant GBMF4541 to PJH. This work made use of the Materials Research Science and Engineering Center Shared Experimental Facilities supported by the National Science Foundation (NSF) (Grant No. DMR-0819762). The WTe$_2$ crystal growth performed at Princeton University was supported by an NSF MRSEC grant, DMR-1420541 (QDG and RJC).  KW and TT acknowledge support from the Elemental Strategy Initiative conducted by the MEXT, Japan and JSPS KAKENHI Grant Numbers JP15K21722. HS and LF were supported by the U.S. Department of Energy Office of Basic Energy Sciences, Division of Materials Sciences and Engineering under award DE-SC0010526. TRC was supported by the Ministry of Science and Technology and National Cheng Kung University, Taiwan, and also acknowledges National Center for Theoretical Sciences (NCTS), Taiwan for technical support. 
 
\textbf{Author contributions:}  QM and SYX performed the measurements and analysed the data with help from AMMV. VF and SW fabricated the devices. QDG and RJC grew the bulk WTe$_2$ single crystals. KW and TT grew the bulk hBN single crystals. TRC, GC and HL calculated the first-principles band structures. SYX, QM, HS and LF did theoretical analysis of the Berry curvature dipole with help from CKC. HS and LF performed analysis of the $k\cdot{p}$ model. SYX and QM wrote the manuscript with input from all authors. PJH and NG supervised the project. 

\textbf{Competing financial interests:} The authors declare no competing financial interests.

\clearpage

\addtocounter{figure}{-1}
\begin{figure*}[t!]
\caption{ \textbf{c,d} Same as panels (a,b) but for the photocurrent along $\hat{a}$ ($I_{\hat{a}}$). The dots in panels (a,c) show the position chosen for the polarization-dependent data in panels (b,d). \textbf{e,f,} $I_{\hat{a}}$ as a function of laser spot (panel (e)) along the dotted line in panel (c) or as a function of polarization (panel (f)) for three different doping levels. \textbf{g,h} Same as panels (e,f) but for three different displacement fields. \textbf{i,j} We color the top and bottom atomic layers differently as the out-of-plane $\vec{D}$ causes the two layers to have different on-site potential energies. In bilayer graphene (panel (j)) and also MoS$_2$, an out of plane $\vec{D}$ field only causes an out-of-plane (+$\hat{c}$ to $-\hat{c}$) polarity. By contrast, because the monolayer WTe$_2$ lattice features a tilted parallelogram (panel (i)), an out of plane $\vec{D}$ field can give rise to an in-plane polarity along $\hat{b}$.}
\end{figure*}

\clearpage
\begin{figure*}[t]
\includegraphics[width=15cm]{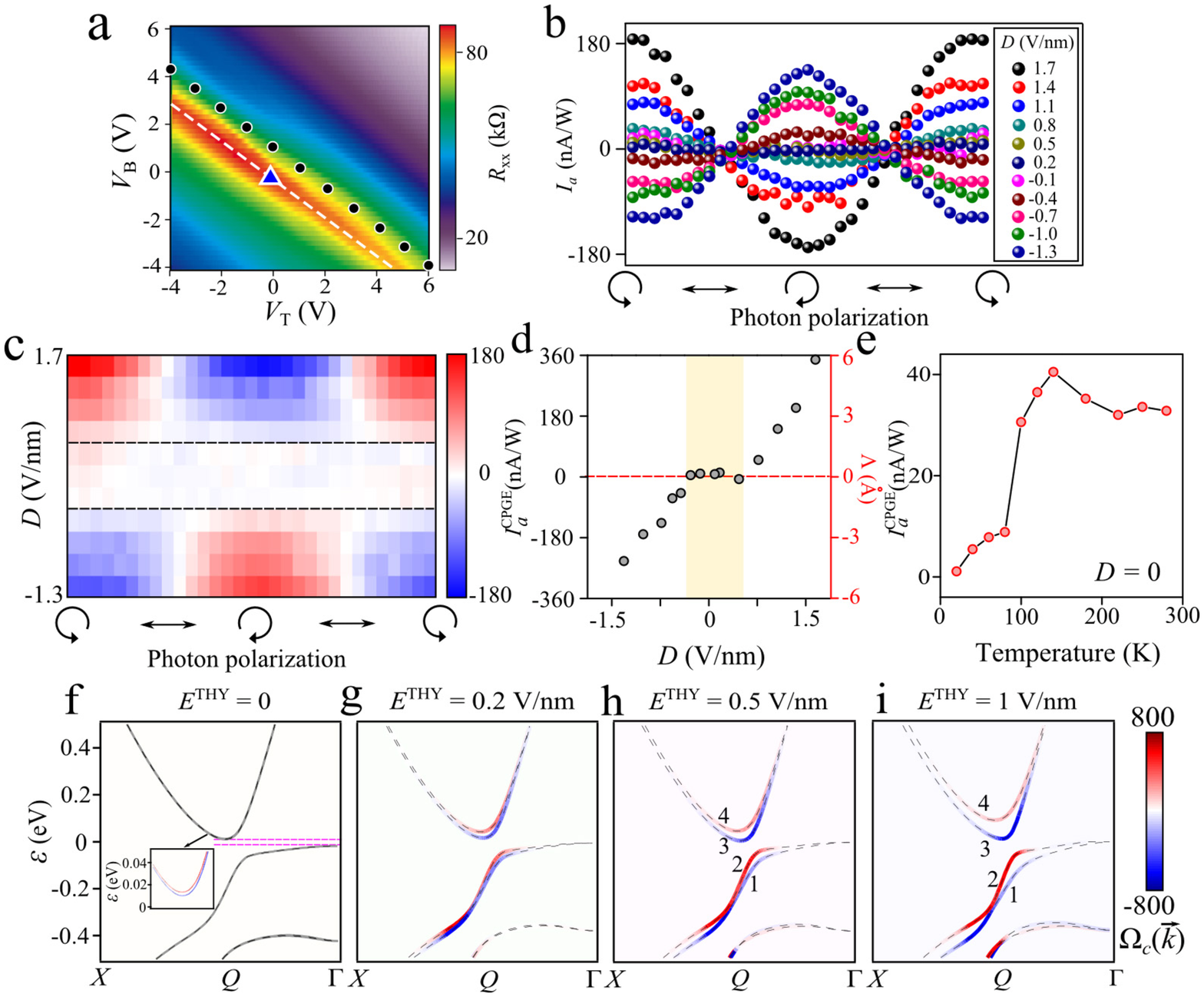}
\caption{{\bf Systematic control of the circular photogalvanic current by displacement fields.} \textbf{a,} Longitudinal DC resistance ($R_{xx}$) as a function of the top and bottom gate voltages at $T=20$ K. The dotted line defines the direction along which one can vary the displacement field while keeping the charge density invariant. \textbf{b,c,} Polarization dependent CPGE currents for different displacement fields at $T=20$ K. \textbf{d,} Left-vertical axis shows the $I_{\hat{a}}^{\textrm{CPGE}}=I_{\hat{a}}(\textrm{RCP})-I_{\hat{a}}(\textrm{LCP})$ as a function of the displacement field at $T=20$ K. Right-vertical axis shows our estimate of the corresponding Berry curvature dipole for each data point (see SI. IV for more details). \textbf{e,} Temperature dependent $I_{\hat{a}}^{\textrm{CPGE}}$ in the absence of displacement fields. \textbf{f-i,} First-principles calculated band structure and Berry curvature (represented by the blue-red color) of monolayer WTe$_2$ using the Heyd-Scuseria-Ernzerhof (HSE) method \cite{heyd2003hybrid} for different out-of-plane electric fields. The four low-energy bands are labeled as 1-4. By fixing the hybrid parameter at HSE$=0.4$ (Figs.~\ref{Fig3}f-i), we obtain a global band gap around $\sim30$ meV, which leads to a minimum direct band gap of $\sim150$ meV without the displacement field. The minimum direct band gap decreases back to $\leq120$ meV for $E\geq0.2$ V/nm (panel (g)) . The electric field ($\vec{E}^{\textrm{THY}}$) here can be related to the displacement field by $\vec{D}^{\textrm{THY}}=\epsilon^{\textrm{WTe}_2}\vec{E}^{\textrm{THY}}$.}
\label{Fig3}
\end{figure*}

\clearpage
\begin{figure*}[t]
\includegraphics[width=17cm]{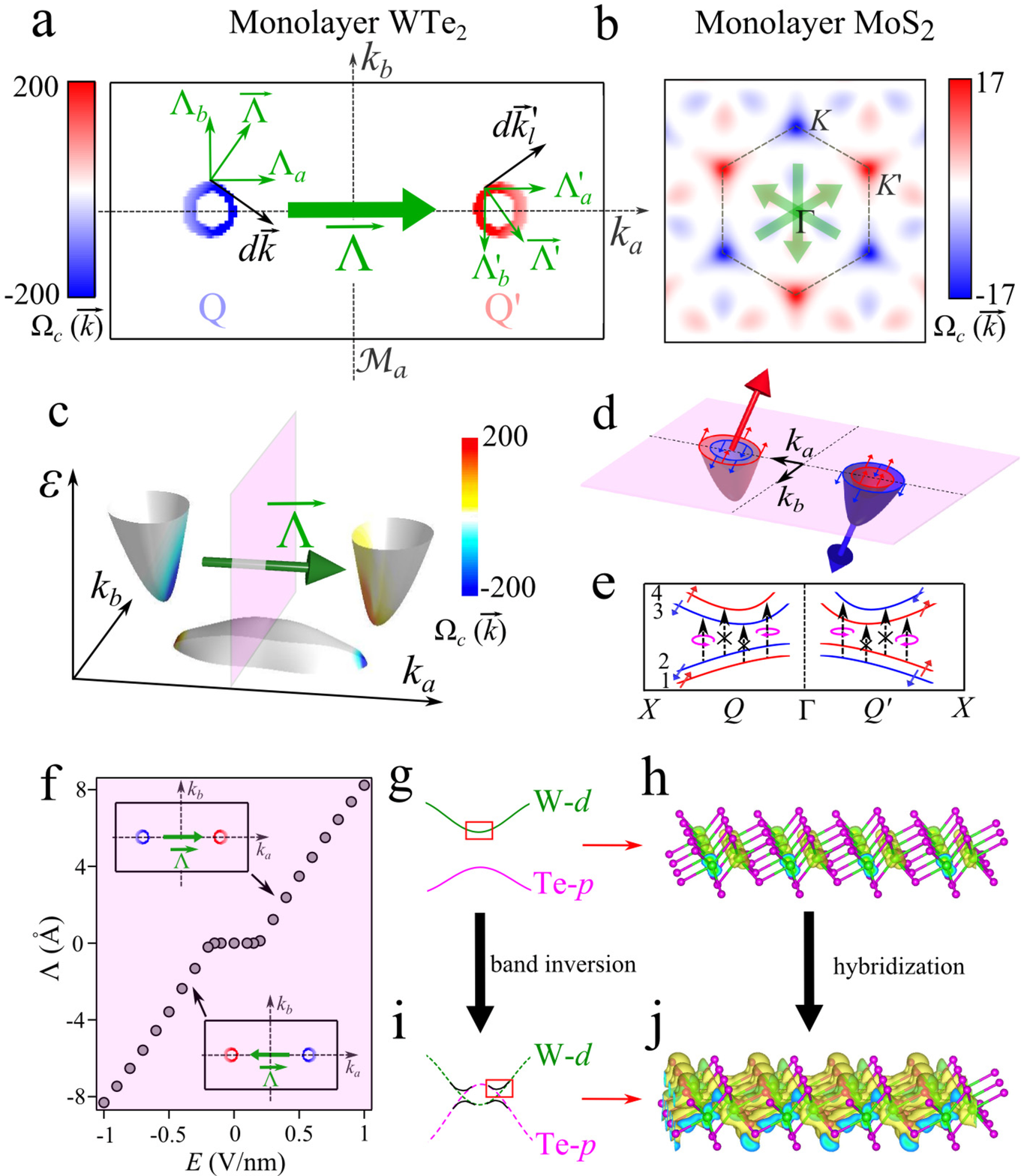}
\caption{ {\bf Berry curvature dipole and its control via a topological field effect.} \textbf{a,c} Berry curvature ($\Omega_c(\vec{k})$) of monolayer WTe$_2$ as a function of energy and momentum with the displacement field set at $E=+0.5$ V/nm (same as Fig. 3i). \textbf{a,} $\Omega_c(\vec{k}$) of band 3 along the $k$-contours that correspond to an $\hbar\omega=120$ meV inter-band transition.}
\label{Fig4}
\end{figure*}
\addtocounter{figure}{-1}
\begin{figure*}[t!]
\caption{Note that $k$ points that allow a fixed $\hbar\omega$ inter-band transition must be a closed contour because one can think of it as the constant energy contour of a new band, whose energy is defined as the energy difference between bands 3 and 2 ($\Delta{\varepsilon}(\vec{k})=\varepsilon_{3}(\vec{k})-\varepsilon_{2}(\vec{k})$). The small arrows depict the Berry curvature dipole contributed from two infinitesimal segments ($d\vec{k}, d\vec{k}'$) on the two contours. The big arrow shows the total Berry curvature dipole integrated over the contours (see Eq.~\ref{Dipole}). \textbf{b,} Berry curvature of monolayer MoS$_2$ as a comparison. \textbf{d,} The canted Zeeman-like spin texture of the lowest two conduction bands (bands 3 and 4) on the constant energy contours. \textbf{e,} Spin direction and optical selection rules between the lowest conductions and valence bands (1-4). \textbf{f,} Calculated BC dipole as a function of the electric displacement field. \textbf{i,j} Calculated real-space distribution of the wavefunction amplitude at the inverted gap edge ($Q$ point) in the presence of the band inversion. \textbf{g,h} Same as panels (i,j) without the band inversion.}
\end{figure*}

\newpage
\clearpage

\vspace{3cm}
\begin{center}
Supplemental Information for 
\end{center}
\vspace{0.45cm}
\textbf{
\begin{center}
{This file includes:\\}
\end{center}
}
\vspace{1cm}

\begin{tabular}{l l}
\underline{I.} & Symmetry analysis of our CPGE data  \\
\underline{II.} &  Spin polarizations of monolayer WTe$_2$ \\
\underline{III.} & Derivation for the Berry curvature dipole and CPGE \\
\underline{IV.} & Experimental estimation of the Berry curvature dipole \\
\underline{V.} & Supplementary discussions \\
\enspace  V.1. & The bulk and monolayer structures of WTe$_2$\\
\enspace  V.2. & Symmetry allowed CPGE in 2D materials \\

\end{tabular}

\date{\today}

\small

\setcounter{figure}{0}
\renewcommand{\figurename}{\textbf{Supplementary Figure}}

\setcounter{equation}{0}

\section*{I. Symmetry analysis of our CPGE data}

We present a symmetry analysis of the observed CPGE. In the absence of a displacement field, the observed CPGE is described by the second-order photocurrent susceptibility tensor $\chi_{ijk}^{(2)}$ \cite{ma2017direct, de2017quantized}:

\begin{eqnarray}
J_i^{\textrm{CPGE}} &=& \chi_{ijk}^{(2)}E_j(\omega)E_k^*(\omega),
\label{X2}
\end{eqnarray} where $J^{\textrm{CPGE}}$ is the CPGE current, $E(\omega)$ and $\omega$ are the electric field and frequency of light, and the tensor indices $i,j,k$ span the sample coordinates ${a,b,c}$. All tensor components identically vanish in the presence of inversion symmetry. When inversion symmetry is broken, one needs to further investigate the role of other crystalline symmetries. Specifically, relevant to our experiments are the in-plane CPGE currents along $a$ and $b$ with normal incident light. They are described by $\chi_{aab}^{(2)}$ and $\chi_{bab}^{(2)}$ respectively. Here we only consider the $1T_d$ phase because the inversion symmetric $1T'$ phase doesn't allow any CPGE. The mirror symmetry $\mathcal{M}_a$ in the $1T_d$ structure forces any component with an odd number of $a$ to vanish. Thus we have $J_a^{\textrm{CPGE}}=\chi_{aab}^{(2)} E_a(\omega)E_b^*(\omega)\neq0$ and $J_b^{\textrm{CPGE}}=\chi_{bab}^{(2)}E_a(\omega)E_b^*(\omega)=0$, which is consistent with our directional dependence of the CPGE data in the main text. Therefore, our data (Figs. 2b,d in the main text) and the above symmetry analysis collectively determine the crystalline directions $\hat{a}$ and $\hat{b}$.

We have intentionally designed the Hall bar contacts to be roughly aligned with the crystalline axes during the fabrication stage. Before putting electrical contacts on an obtained monolayer crystal, we first estimated its crystalline orientation by the surrounding thicker flakes. The thicker flakes, similar to the bulk crystals, usually had a rectangular shape, where the long and short axes correspond to the crystalline $\hat{a}$ and $\hat{b}$ directions \cite{ali2014large}. Since the orientation of surrounding thicker flakes are roughly aligned, we thus assumed that the monolayer crystal has the same orientation as its surroundings and put the electric contacts accordingly. The further, independent assignment of the crystalline orientations are done by combining the directional dependent CPGE data with the symmetry analysis as shown above. 

In order to also include the displacement field $\vec{D}$ effect, we rewrite the CPGE as 

\begin{align}
J_a^{\textrm{CPGE}} &=  \chi_{aab}^{(2)}E_a(\omega)E_b^*(\omega)+\chi_{aabc}^{(3)}E_a(\omega)E_b^*(\omega)D_c \nonumber\\
&= [\chi_{aab}^{(2)}+\chi_{aabc}^{(3)}D_c]E_a(\omega)E_b^*(\omega) = \widetilde{\chi}_{aab}^{(2)}(D_c)E_a(\omega)E_b^*(\omega),
\label{X3}
\end{align} where $\chi_{aab}^{(2)}$ (same as in Eq.~\ref{X2}) is the second order photocurrent susceptibility tensor, $\chi_{aabc}^{(3)}$ is a third order photocurrent susceptibility tensor, and $D_c$ is the static displacement field. Under this construction, $\chi_{aab}^{(2)}$ can account for the CPGE current in the absence of the $\vec{D}$, which is due to the intrinsic inversion symmetry breaking of the WTe$_2$ monolayer crystal, whereas $\chi_{aabc}^{(3)}$ gives rise to the CPGE from $\vec{D}$. Because of the weak (intrinsic) inversion-breaking of monolayer WTe$_2$, a sufficiently large $\vec{D}$ can dominate and cause a sign-reversal of the CPGE. Because the displacement field is always along $\hat{c}$, we can also redefine the second order tensor to cover the displacement field effect by writing $\widetilde{\chi}_{aab}^{(2)}(D_c)=\chi_{aab}^{(2)}+\chi_{aabc}^{(3)}D_c$. 

\section*{II. Spin polarizations of monolayer WTe$_2$}

\begin{figure*}[h]
\centering
  \begin{minipage}[c]{0.75\textwidth}
\includegraphics[width=12cm]{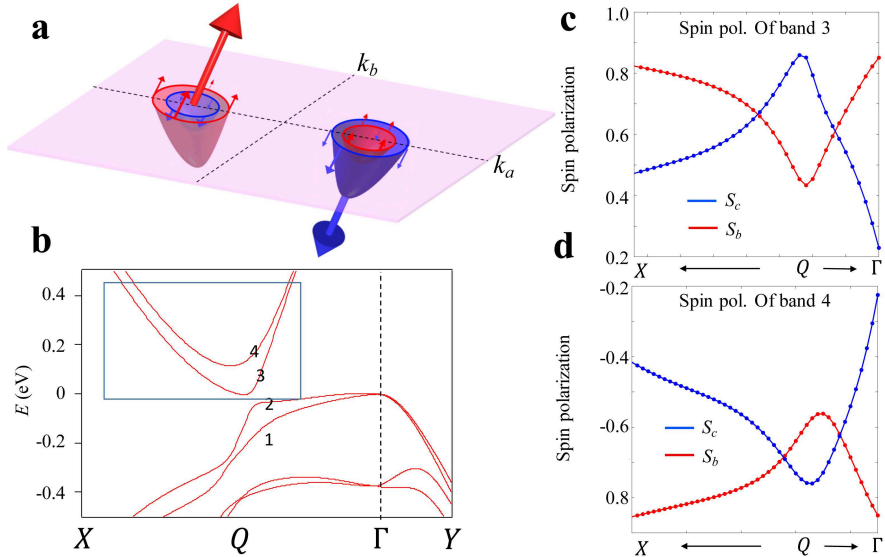}
  \end{minipage}\hfill
  \begin{minipage}[c]{0.25\textwidth}
\caption{\textbf{Spin polarizations of monolayer WTe$_2$} \textbf{a,} Schematic illustration of the spin textures on the constant energy contours of the conduction bands. \textbf{b,} Band structure with an $E=1$ V/nm displacement field. \textbf{c,d} The $\hat{b}$ and $\hat{c}$ components of the spin polarization ($S_b, S_c$) of bands 3 and 4. The $k$ range for these calculations are indicated by the blue box in panel (b). }
  \end{minipage}
\label{WTe2_spin_cal}
\end{figure*}

\section*{III. Derivation for the Berry curvature dipole and CPGE}

The CPGE can be mathematically described as:

\begin{align}
\vec{J}^{\textrm{CPGE}} = & - \frac{2\pi{e}\tau}{\hbar}\sum_{I,F}\int\frac{d^2k}{(2\pi)^2}[\Delta\vec{v}(\vec{k})]\mathcal{V}(\vec{k})\delta(\Delta\mathlarger{\varepsilon}(\vec{k})-\hbar \omega)[\Delta{f}(\mu, \vec{k})].
\label{CPGE}
\end{align}
In this equation, $\vec{J}^{\textrm{CPGE}}$ is the difference between the photocurrents due to RCP and LCP light, $\tau$ is the relaxation time, $\mathcal{V}(\vec{k})=\arrowvert{\mathcal{P}^{\textrm{RCP}}(\vec{k})}\arrowvert^2-\arrowvert{\mathcal{P}^{\textrm{LCP}}(\vec{k})}\arrowvert^2$ describes the difference between the optical transition probability for RCP and LCP light, $I,F$ are all possible initial and final states that satisfy the energy conservation enforced by the $\delta$ function, and $\Delta\vec{v}$, $\Delta\mathlarger{\varepsilon}(\vec{k})$ $\Delta{f}(\mu, \vec{k})$ are the difference of the group velocity, energy, and Fermi-Dirac distribution between the initial and final states. $\mathcal{P}$, the optical transition dipole, is defined as

\begin{align}
\mathcal{P} = & \frac{e}{m_e}\langle{F}\arrowvert \vec{A}\cdot\vec{p}\arrowvert{I}\rangle,
\label{Transition_dipole}
\end{align}

where $e$ and $m_e$ are the charge and mass of a bare electron, $\arrowvert{I}\rangle$ and $\arrowvert{F}\rangle$ are the wavefunctions of the initial and final states, and $\vec{A}$ is the vector potential of light and $\vec{p}$ is the momentum operator. The momentum operator $\vec{p}$ is defined as

\begin{align}
\vec{p}=\frac{m_e}{i\hbar}[\vec{r}, H],
\label{Momentum}
\end{align}

Using Eq.~\ref{Transition_dipole}, $\mathcal{V}(\vec{k})$ under a normal incident, circularly polarized light can be expressed as

\begin{align}
\mathcal{V}(\vec{k}) & =\arrowvert{\mathcal{P}^{\textrm{RCP}}(\vec{k})}\arrowvert^2-\arrowvert{\mathcal{P}^{\textrm{LCP}}(\vec{k})}\arrowvert^2 \nonumber\\
&=(\frac{Ae}{m_e})^2\bigg[\Big\arrowvert\langle{F}\arrowvert({p_x}+i{p_y})\arrowvert{I}\rangle\Big\arrowvert^2-\Big\arrowvert\langle{F}\arrowvert({p_x}-i{p_y})\arrowvert{I}\rangle\Big\arrowvert^2\bigg] \nonumber\\
&= \frac{2A^2e^2}{m_e^2}\bigg[i\langle{F}\arrowvert{p_y}\arrowvert{I}\rangle\langle{I}\arrowvert{p_x}\arrowvert{F}\rangle-i\langle{F}\arrowvert{p_x}\arrowvert{I}\rangle\langle{I}\arrowvert{p_y}\arrowvert{F}\rangle\bigg]
\label{ME}
\end{align}
where $\arrowvert{I}\rangle$ and $\arrowvert{F}\rangle$ are the Bloch wavefunction of the possible initial and final states for the inter-band transition, ${p_x}$ and ${p_y}$ are the momentum operators, and $A$ is the vector potential of light. 

The BC for an $N$-band system (in 2D, the BC is only defined along the out-of-plane direction) is defined as

\begin{align}
\Omega_z^{\textrm{band n}}(\vec{k})=i\sum_{n'\neq{n}}\frac{\langle{n'}\arrowvert{\frac{\partial H}{\partial k_x}}\arrowvert{n}\rangle\langle{n}\arrowvert {\frac{\partial H}{\partial k_y}} \arrowvert{n'}\rangle-\langle{n'}\arrowvert {\frac{\partial H}{\partial k_y}} \arrowvert{n}\rangle\langle{n} {\frac{\partial H}{\partial k_x}} \arrowvert{n'}\rangle}{(\mathlarger{\varepsilon}_{n'}-\mathlarger{\varepsilon}_{n})^2}.
\label{BC1}
\end{align}

Equations~\ref{CPGE}-\ref{BC1} are applicable to the general case. We see that, in general, the CPGE is a complicated process: When the photon energy $\hbar\omega$ is large, the interband transition involves many bands. Then one needs to sum over all possible initial and final states that satisfy the energy conservation ($\sum_{I,F}$ in Eq.~\ref{CPGE}). 

In a two-band system, a number of theoretical works \cite{xiao2007valley, yao2008valley, wu2013electrical, de2017quantized} have shown that $\mathcal{V}(\vec{k})$ can be directly expressed by the Berry curvature:

\begin{align}
\mathcal{V}(\vec{k}) &=\frac{2e^2}{\hbar^2}\,A^2\,\Omega_z^{\textrm{CB}}(\vec{k})\,(\Delta\mathlarger{\varepsilon})^2.
\label{ME2_0}
\end{align} Here we repeat the derivation of Eq.~\ref{ME2_0}: First, using the Peierls substitution, the momentum operator $p_i$ in the optical transition dipole (Eq.~\ref{ME}) can be calculated by $\frac{m_e}{h}\partial H /\partial {k_i}$ \cite{dresselhaus1967fourier, yao2008valley, xiao2007valley, smith1974photoemission, de2017quantized, photocurrent, taguchi2016photovoltaic,kim2017shift}. Second, for a two-band system, the BC can be simplified as

\begin{align}
\Omega_z^{\textrm{VB}}(\vec{k})=-\Omega_z^{\textrm{CB}}(\vec{k})=i\frac{\langle{v}\arrowvert{\frac{\partial H}{\partial k_x}}\arrowvert{c}\rangle\langle{c}\arrowvert{\frac{\partial H}{\partial k_x}}\arrowvert{v}\rangle-\langle{v}\arrowvert{\frac{\partial H}{\partial k_x}}\arrowvert{c}\rangle\langle{c}\arrowvert{\frac{\partial H}{\partial k_x}}\arrowvert{v}\rangle}{(\Delta\mathlarger{\varepsilon})^2},
\label{BC2}
\end{align}
where $\arrowvert{v}\rangle$ and $\arrowvert{c}\rangle$ are the wavefunctions of the only valence band and the only conduction band in a two-band system. We compare the $N$-band BC (Eq.~\ref{BC1}) and the two-band BC (Eq.~\ref{BC2}). In an $N$-band system, the BC of the $n^{\textrm{th}}$ band $\Omega_z^n(\vec{k})$ is contributed by all the other $N-1$ bands (Eq.~\ref{BC1}). By contrast, in a two-band system, the BC of the valence band $\Omega_z^{\textrm{VB}}(\vec{k})$ is purely contributed by the conduction band and vice versa (Eq.~\ref{BC2}). Under the two-band approximation and using the $p_i\rightarrow\frac{m_e}{h}\partial H /\partial {k_i}$ substitution, $\mathcal{V}(\vec{k})$ can be expressed in terms of the BC:

\begin{align}
\mathcal{V}(\vec{k}) &=\frac{2e^2}{\hbar^2}\,A^2\,\Omega_z^{\textrm{CB}}(\vec{k})\,(\Delta\mathlarger{\varepsilon})^2.
\label{ME2}
\end{align}

From Eq.~\ref{ME2}, one can see that optical transition probability is directly proportional to the BC. States with positive (negative) BC selectively absorbs RCP (LCP) light. This result has been reported in Refs. \cite{xiao2007valley, yao2008valley, wu2013electrical, de2017quantized}.

We now show how the photocurrent (Eq.~\ref{CPGE}) can be connected to the BC dipole

\begin{align}
J^{\textrm{CPGE}}_x = & - \frac{2\pi{e}\tau}{\hbar}\sum_{I,F}\int\frac{d^2k}{(2\pi)^2}[\Delta\vec{v}(\vec{k})]\mathcal{V}(\vec{k})\delta(\Delta\mathlarger{\varepsilon}(\vec{k})-\hbar \omega)[\Delta{f}(\mu, \vec{k})] \nonumber\\
& = - \frac{e^3\tau{A}^2}{\pi\hbar^3}{\int}dk_xdk_y\frac{d(\Delta{\mathlarger{\varepsilon}})}{\hbar{dk_x}}\Omega(\vec{k})(\Delta\mathlarger{\varepsilon})^2\delta(\Delta\mathlarger{\varepsilon}-\hbar\omega) \nonumber\\
&=- \frac{e^3\tau{A}^2}{\pi\hbar^4}{\int}dk_yd(\Delta{\mathlarger{\varepsilon}})\Omega(\vec{k})(\Delta\mathlarger{\varepsilon})^2\delta(\Delta\mathlarger{\varepsilon}-\hbar\omega)\nonumber\\
&=- \frac{e^3\tau{E}^2}{\pi\hbar^2}{\oint}dk_y\Omega(\vec{k}),
\label{CPGE2}
\end{align}
where the closed loop integral ($\oint dk_y$) is defined along the $k$-contours that correspond to an $\hbar \omega$ inter-band transition. In this derivation, we have used Eq.~\ref{ME2} and the definition of group velocity $\Delta\vec{v}(\vec{k})=\frac{1}{\hbar}\nabla_{\vec{k}}(\Delta{\mathlarger{\varepsilon}})$, and dropped the summation ($\sum_{I,F}$) because, in a two-band system, the only initial and final states are the conduction and valence bands. Because the BC $\Omega(\vec{k})$ in 2D only has the out of the plane component, we can rewrite the CPGE into

\begin{align}
\vec{J}^{\textrm{CPGE}}&=\frac{e^3\tau}{\pi\hbar^2}\,\mathcal{I}m\bigg[\vec{E}(-\omega)\times\hat{c}\,\,(\vec{\mathlarger{\mathlarger{\Lambda}}}^{\Omega}\cdot\vec{E}(\omega))\bigg],
\label{CPGE_Dipole}
\end{align}
where $\vec{\mathlarger{\mathlarger{\Lambda}}}^{\Omega}=\oint d\vec{k} \times \vec{\Omega}(\vec{k})$ is the BC dipole, $\vec{E}(\omega)=(E_xe^{i\omega{t}},E_ye^{i(\omega{t}+\gamma)},0)$ describes electric field of normal incident light with a generic polarization. $E_x, E_y$ are positive, real numbers, and $\gamma$ is the phase difference between the $x$ and $y$ components. $\gamma=\pm\frac{\pi}{2}$ and $E_x=E_y$ correspond to the normal incident R(L)CP light. 

We make the following comments in connection to the above derivation: 

\textbf{1. Two-band approximation:} The two-band approximation allows us to simplify things in two aspects. First, the summation over all possible initial and final states ($\sum_{I,F}$) in Eq.~\ref{CPGE} can be dropped because the only initial and final states are the valence and conduction bands, respectively. Second, the BC is simplified: the BC of the valence band $\Omega_z^{\textrm{VB}}(\vec{k})$ is purely contributed by the conduction band and vice versa (Eq.~\ref{BC2}). In our experiments on monolayer WTe$_2$, because the low photon energy $\hbar\omega=120$ meV matches the direct band gap near $Q(Q')$, the optical transition indeed only involves the lowest conduction and the highest valence bands. Below, we further show that the BC of the lowest conduction band is almost entirely contributed from the highest valence band and vice versa. 

\begin{figure*}[h]
\centering
\includegraphics[width=14cm]{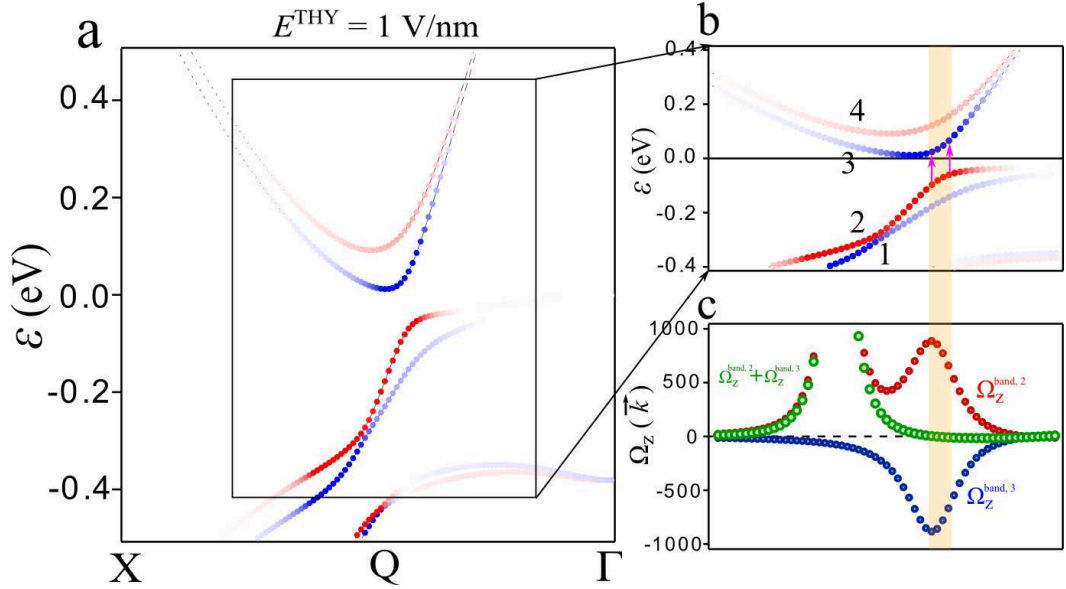}
\caption{\textbf{Berry curvatures of the low-energy electron states of monolayer WTe$_2$.} \textbf{a,b,} Band structure and BC of monolayer WTe$_2$ with $E^{\textrm{THY}}=0.1$ eV/nm. The lowest conduction and highest valence bands are labeled as bands 1-4 following the same convention as the main text. \textbf{c,} BCs of band 2 ($\Omega_z^{\textrm{band 2}}(\vec{k})$), band 3 ($\Omega_z^{\textrm{band 3}}(\vec{k})$) and their sum ($\Omega_z^{\textrm{band 3}}(\vec{k})+\Omega_z^{\textrm{band 2}}(\vec{k})$). The $k$ region corresponding to the inverted gap edge is highlighted by the orange shaded area. The $\hbar\omega=120$ meV optical transition in our experiments are indicated by the pink arrows.}
\label{BBfigure}
\end{figure*}

Following the convention used in the main text, we label the lowest four bands as bands 1-4. We zoom in near the inverted band gap edge and show in Fig.~\ref{BBfigure}c the Berry curvatures of the lowest conduction band (band 3) $\Omega_z^{\textrm{band 3}}(\vec{k})$ and highest valence band (band 2) $\Omega_z^{\textrm{band 2}}(\vec{k})$ as well as their sum ($\Omega_z^{\textrm{band 3}}(\vec{k})+\Omega_z^{\textrm{band 2}}(\vec{k})$). At the inverted gap edge, the Berry curvatures of the lowest conduction and highest valence bands are nearly equal but opposite. \textbf{Therefore, for the inter-band transition considered in our experiment (indicated by the pink arrows in Fig.~\ref{BBfigure}b), the two-band system is a very good approximation. } This is consistent with the $k\cdot{p}$ results in section II.

On the other hand, as one moves away from the inverted gap edge, $\Omega_z^{\textrm{band 2}}(\vec{k})$ shows a large peak, which is absent in $\Omega_z^{\textrm{band 3}}(\vec{k})$ (Fig.~\ref{BBfigure}c). This peak $\Omega_z^{\textrm{band 2}}(\vec{k})$ is mainly contributed from band 1 (see Fig.~\ref{BBfigure}b), which energetically approaches band 2 in that $k$ region. This contribution is not seen by the $k\cdot{p}$ results in section II, where $\Omega_z^{\textrm{band 2}}(\vec{k})=-\Omega_z^{\textrm{band 3}}(\vec{k})$ always holds. Thus it comes from additional details of monolayer WTe$_2$, which is not captured by the main physical picture presented in section II.

\textbf{2. The ``two-band'' CPGE under oblique incidence:} We have shown that the ``two-band'' CPGE under normal incidence uniquely measures the BC dipole. A natural question is how the situation changes with oblique incidence. Here we show that ``two-band'' CPGE under oblique incidence picks up an additional contribution that is not due to the BC and BC dipole. Specifically, this additional contribution comes from an atomic scale contribution along the out-of-plane $\hat{z}$ direction.

Without losing generosity, we assume that the light is tilted away $\hat{c}$ within the $\hat{a}-\hat{c}$ plane with an angle $\theta$. In this case, $\mathcal{V}(\vec{k})$ becomes 

\begin{align}
\mathcal{V}(\vec{k}) & =\arrowvert{\mathcal{P}^{\textrm{RCP}}(\vec{k})}\arrowvert^2-\arrowvert{\mathcal{P}^{\textrm{LCP}}(\vec{k})}\arrowvert^2 \nonumber\\
&=(\frac{Ae}{m_e})^2\bigg[\Big\arrowvert\langle{c}\arrowvert({p_x}\cos\theta+{p_z}\sin\theta+i{p_y})\arrowvert{v}\rangle\Big\arrowvert^2-\Big\arrowvert\langle{v}\arrowvert({p_x}\cos\theta+{p_z}\sin\theta-i{p_y})\arrowvert{c}\rangle\Big\arrowvert^2\bigg] \nonumber\\
&\qquad\quad\,\,\,\,+ \bigg[i\langle{c}\arrowvert{p_y}\arrowvert{v}\rangle\langle{v}\arrowvert{p_z}\arrowvert{c}\rangle-i\langle{c}\arrowvert{p_z}\arrowvert{v}\rangle\langle{v}\arrowvert{p_y}\arrowvert{c}\rangle\bigg]\sin\theta\Bigg) \nonumber\\
&= \frac{2A^2e^2}{\hbar^2}\Bigg(\Omega_z^{\textrm{CB}}(\vec{k})\,(\Delta\mathlarger{\varepsilon})^2\cos\theta + \bigg[\langle{c}\arrowvert{\frac{\partial H}{\partial k_y}}\arrowvert{v}\rangle\langle{v}\arrowvert{[z, H]}\arrowvert{c}\rangle-\langle{c}\arrowvert{[z, H]}\arrowvert{v}\rangle\langle{v}\arrowvert{\frac{\partial H}{\partial k_y}}\arrowvert{c}\rangle\bigg]\sin\theta \Bigg)\nonumber\\
&= \frac{2A^2e^2}{\hbar^2}\Bigg(\Omega_z^{\textrm{CB}}(\vec{k})\,(\Delta\mathlarger{\varepsilon})^2\cos\theta + \bigg[\langle{c}\arrowvert{\frac{\partial H}{\partial k_y}}\arrowvert{v}\rangle\langle{v}\arrowvert{z}\arrowvert{c}\rangle-\langle{c}\arrowvert{z}\arrowvert{v}\rangle\langle{v}\arrowvert{\frac{\partial H}{\partial k_y}}\arrowvert{c}\rangle\bigg]\Delta\mathlarger{\varepsilon}\sin\theta \Bigg).
\label{ME_tilt}
\end{align}

The key point in this derivation is that, in a 2D crystal, the ``$p_i\to\frac{m_e}{h}\partial H /\partial {k_i}$'' substitution is only valid for $i=x$ or $i=y$ because the 2D crystal only has translational symmetry inside the plane. On the other hand, translational symmetry is broken along $\hat{z}$ for a 2D crystal ($k_z$ is ill-defined). Therefore, $p_z$ can only be calculated by its definition $p_z=\frac{im_e}{h}[z, H]$. From Eq.~\ref{ME_tilt}, one can see that the first term is the CPGE current from BC dipole (Eq.~\ref{CPGE_Dipole}). On the other hand, the second term is a new contribution because of the tilted light ($\theta\neq0$). Along $\hat{z}$, the wavefunction is strongly localized in the vicinity of the 2D layer. Therefore, $\langle{v}\arrowvert{z}\arrowvert{c}\rangle$ (the second term in Eq.~\ref{ME_tilt}) is an atomic scale contribution along the out-of-plane $\hat{z}$ direction, which is irrelevant to the Berry curvature (see appendix for more details). All $\chi_{xxz}$ and $\chi_{yyz}$ in table~\ref{CPGE_2D_crystals_1} arise from the (non-Berry-curvature) atomic scale contribution along the out-of-plane $\hat{z}$ direction.

\section*{IV. Experimental estimation of the Berry curvature dipole}

In this section, we estimate the Berry curvature dipole based on Eq.~\ref{CPGE_Dipole}. With a circularly polarized light written as $(E_0, \pm i E_0, 0) e^{i\omega t}$ ($E_0$ is a real number) and $I_\textrm{light} = \frac{\alpha P}{A}=cn\epsilon_0 E_0^2$, We then have

\begin{align}
\mathlarger{\Lambda}= \frac{\pi \hbar^2 c \epsilon_0 L \tilde{I}^\textrm{CPGE} }{e^3\tau \alpha},
\label{Dipole}
\end{align}

where $c$ is the speed of light, $\epsilon_0$ is the dielectric constant of vacuum, $\tilde{I}^\textrm{CPGE}$ is the CPGE current as shown in the main text, and $L$, $\tau$, and $\alpha$ are the length, the relaxation time, and the absorption rate of the sample. 

\begin{figure}[h]
  \begin{minipage}[c]{0.6\textwidth}
	\includegraphics[width=8cm]{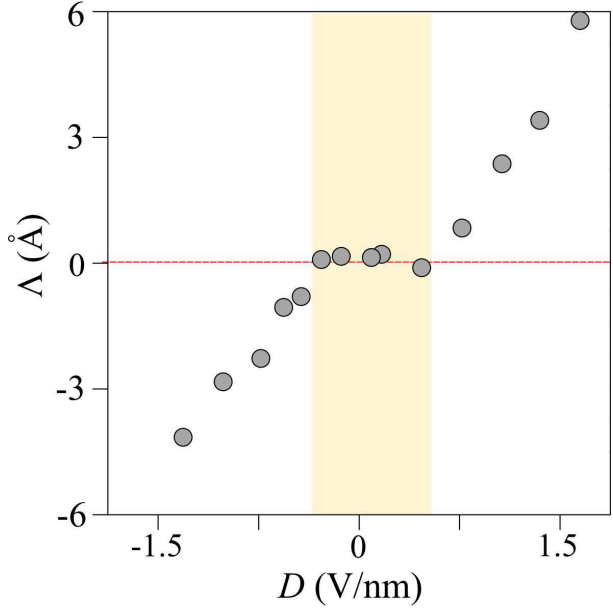}
  \end{minipage}\hfill
  \begin{minipage}[c]{0.4\textwidth}
    \caption{Experimental estimation of the BC dipole $\mathlarger{\Lambda}$ as a function of the external displacement field. The data points are the same as Fig. 3d of the main text.} \label{BC_exp}
  \end{minipage}
\end{figure}

For our sample used in the main text, the length $L \simeq 5 \mu \textrm{m}$; The absorption ratio for a monolayer crystal is roughly $\alpha \approx 0.01$ \cite{fang2013quantum, stauber2015universal, merthe2016transparency}; The relaxation time tao can be estimated from our transport measured electron mobility $\tilde{I}^\textrm{CPGE}$ in Fig. 3d of the main text; The measured CPGE currents ($\tilde{I}^\textrm{CPGE}$) are ploted in Fig. 3d of the main text. By using these numbers, we get $\mathlarger{\Lambda}$ as shown in Fig.~\ref{BC_exp}, which is qualitatively consistent with the calculated BC dipole in Fig. 4f.

\section*{V. Supplementary discussions}
\subsection*{V.1. The bulk and monolayer structures of WTe$_2$} 

In this section, we present a systematic discussion on the bulk and monolayer structures of WTe$_2$ and MoTe$_2$. We refer to the bulk crystal structures as ${bulk-1T'}$ and ${bulk-1T_d}$ and the monolayer structures as ${monolayer-1T'}$ and ${monolayer-T_d}$.

\begin{figure*}[h]
\includegraphics[width=14cm]{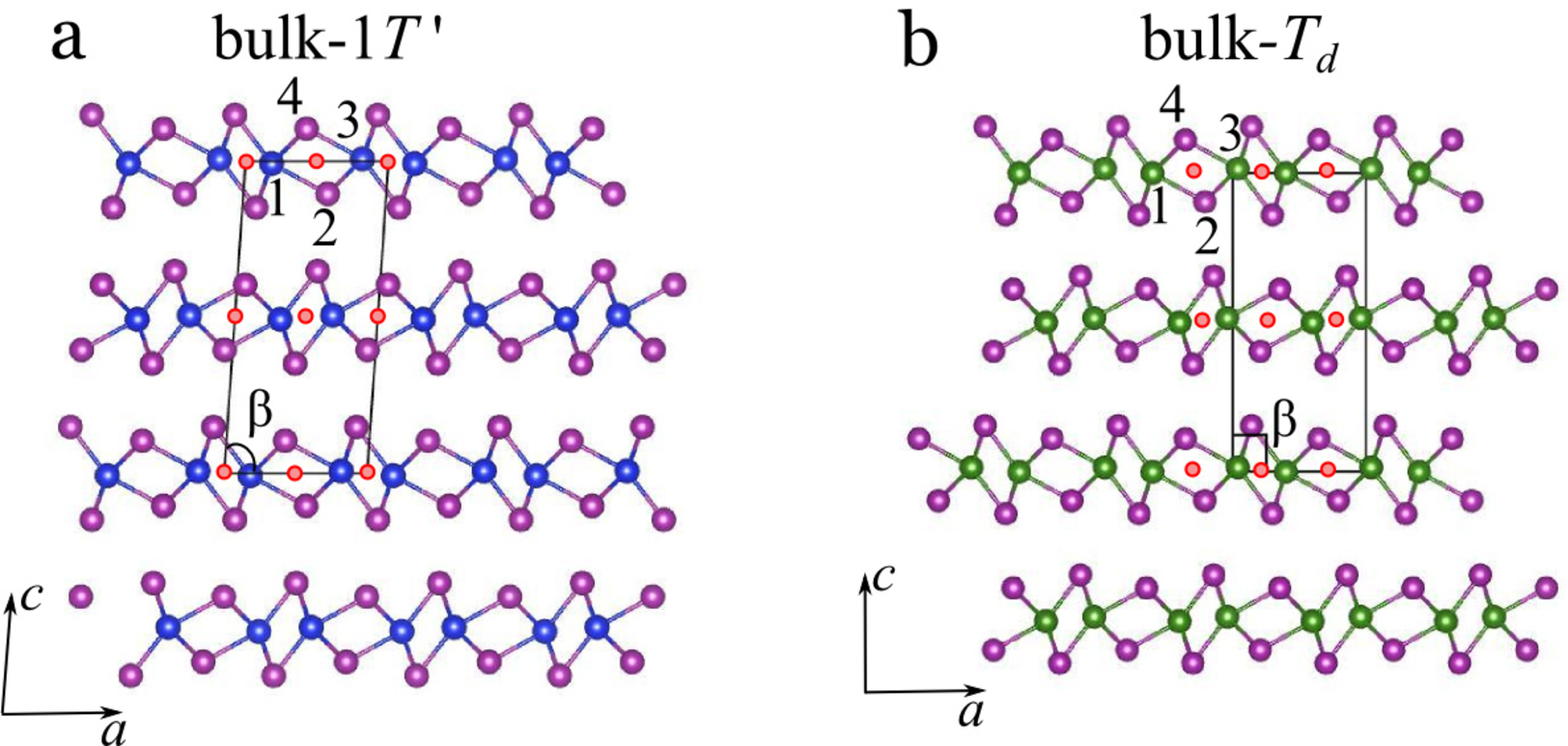}
\caption{\textbf{the $bulk-1T'$ and $bulk-T_d$ structures}}
\label{bulk_1T_Td}
\end{figure*}

$\mathbf{bulk-1T'}$: The ${bulk-1T'}$ phase is a layered, primitive monoclinic structure. Its unit cell is titled ($\beta\neq90^{\circ}$, see Fig.~\ref{bulk_1T_Td}a). This structure has two independent symmetries: a mirror plane $\mathcal{M}_{a}$ and a two-fold rotational axis $\mathcal{C}_{2a}$. The combination of these two symmetries leads to the inversion symmetry. The inversion symmetry and the $\mathcal{C}_{2a}$ rotational symmetry can be visualized in the following way: Each nearby four atoms form a perfect parallelogram (atoms 1-4 in Fig.~\ref{bulk_1T_Td}a), whose center is noted by an orange circle. These orange circles further form larger parallelograms that are similar to the unit cell (the black lines in Fig.~\ref{bulk_1T_Td}a). Therefore, each orange circle represent an inversion center and a $\mathcal{C}_{2a}$ rotational center of the ${bulk-1T'}$ crystal lattice.

$\mathbf{bulk-T_d}$: The ${bulk-T_d}$ is further distorted from the ${bulk-1T'}$ in the sense that its unit cell becomes straight up ($\beta=90^{\circ}$, see Fig.~\ref{bulk_1T_Td}b). Such a distortion breaks the inversion symmetry $\mathcal{I}$ and the in-plane two-fold rotational symmetry $\mathcal{C}_{2a}$. The breaking of inversion ($\mathcal{I}$) and $\mathcal{C}_{2a}$ symmetries can be visualized in the following way: As we change from ${bulk-1T'}$ to ${bulk-T_d}$, the distortion has two consequences: (1) The quadrilateral formed atoms 1-4 (Fig.~\ref{bulk_1T_Td}b) deviates from a perfect parallelogram. The deviation is very small so the quadrilateral still looks like a parallelogram (this deviation is further explained later in the monolayer structures). (2) The centers of these approximate parallelograms (the orange circles in Fig.~\ref{bulk_1T_Td}b) also do not form rectangles that are similar to the unit cell (the black lines in Fig.~\ref{bulk_1T_Td}a). As a result, both inversion ($\mathcal{I}$) and $\mathcal{C}_{2a}$ symmetries are broken. In addition to the above symmetry breakings, the distortion also gives rise to the following new symmetries, the mirror plane $\mathcal{M}_{b}$ and the out-of-plane two-fold rotational symmetry $\mathcal{C}_{2c}$. Therefore, $bulk-T_d$ is an inversion breaking, orthorhombic phase.

\begin{center}
\begin{table}[h]
\begin{tabular}{p{4cm}|p{5cm}|p{5cm}}
\hline
 & $bulk-1T'$ & $bulk-T_d$\\
\hline
Structural phase & primitive monoclinic & primitive orthorhombic\\
Lattice constants & $a\neq{b}\neq{c}$ & $a\neq{b}\neq{c}$\\
Angles & $\alpha=\gamma=90^{\circ}$, $\beta\neq90^{\circ}$ & $\alpha=\beta=\gamma=90^{\circ}$\\
Space group & $P2/m(\#11)$ & $Pmn2_1 (\#31)$\\
Point group & $C_{2h}$ & $C_{2v}$\\
Symmetries & $\mathcal{I}$, $\mathcal{C}_{2a}$, $\mathcal{M}_{a}$ & $\mathcal{M}_{a}$, $\mathcal{M}_{b}$, $\mathcal{C}_{2c}$\\
\hline
\end{tabular}
\caption{\textbf{Key properties of the $bulk-1T'$ and $bulk-T_d$ structures.} $\mathcal{I}$ is the inversion symmetry; $\mathcal{C}_{2}$ is a two-fold rotational symmetry; $\mathcal{M}$ is a mirror plane.}
\label{bulk_table}
\end{table}
\end{center}

We now discuss the crystal structures of bulk WTe$_2$ and MoTe$_2$, both of which have been determined by extensive x-ray studies \cite{obolonchik1972chemical, yanaki1973preparation, brixner1962preparation, brown1966crystal, mar1992metal, agarwal1986growth, canadell1990semimetallic, vellinga1970semiconductor, albert1992preparation, zandt2007quadratic, qi2016superconductivity, zhang2016raman, deng2016experimental}.

\textbf{Bulk WTe$_2$:} Bulk WTe$_2$ crystallizes in the inversion breaking $bulk-T_d$ structure, as consistently found in all x-ray studies \cite{obolonchik1972chemical, yanaki1973preparation, brixner1962preparation, brown1966crystal, mar1992metal}.

\textbf{Bulk MoTe$_2$:} Bulk MoTe$_2$ can crystallize in three different structures, i.e., $bulk-1T'$, $bulk-T_d$ and $bulk-2H$ (same as bulk MoS$_2$, not shown in Fig.~\ref{bulk_1T_Td}) depending on the growth condition \cite{agarwal1986growth, canadell1990semimetallic, vellinga1970semiconductor, albert1992preparation, zandt2007quadratic, qi2016superconductivity, zhang2016raman, deng2016experimental}.
	
\begin{figure*}
\centering
\includegraphics[width=14cm]{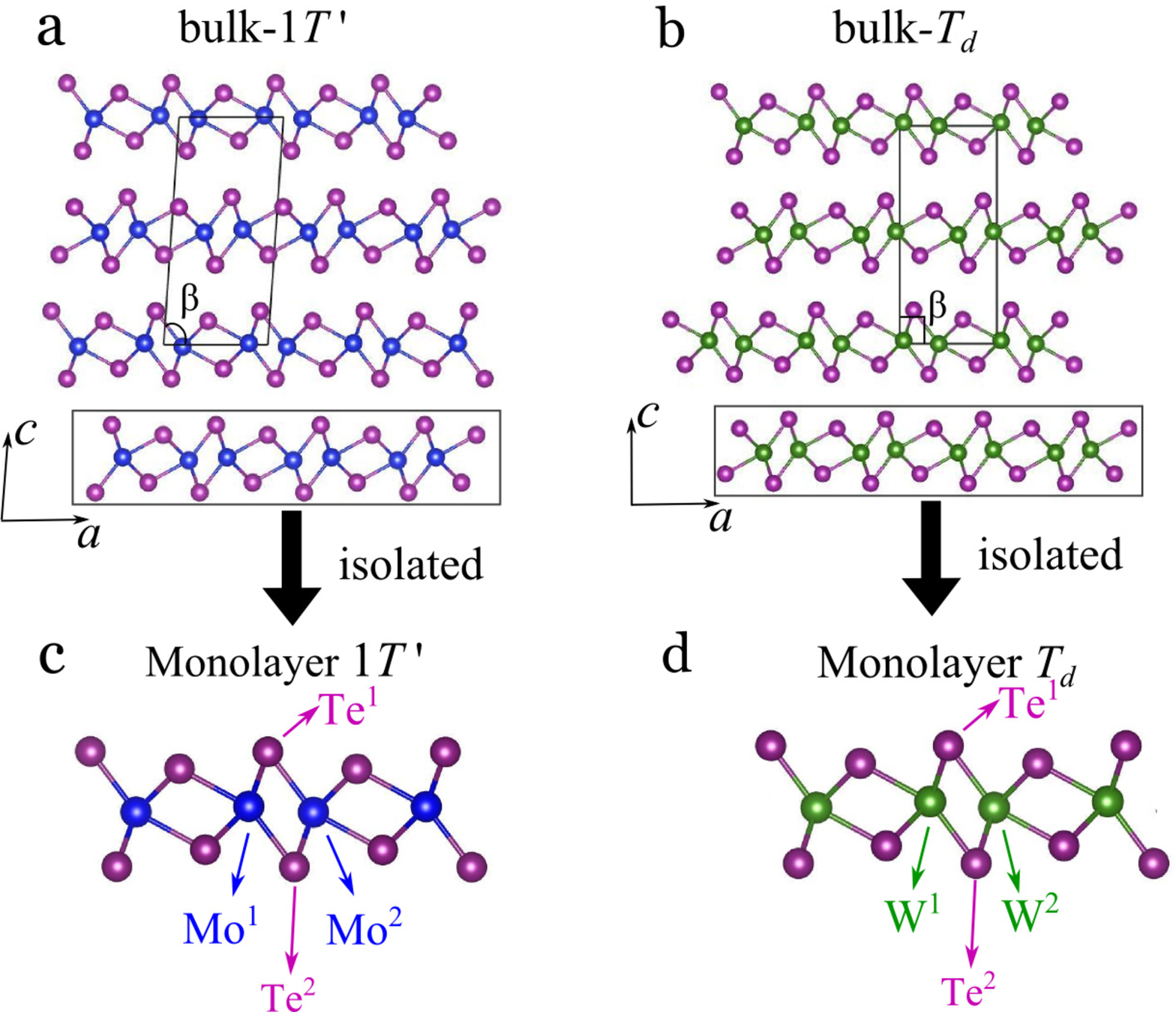}
\caption{\textbf{the $monolayer-1T'$ and $monolayer-T_d$ structures}}
\label{monolayer_1T_Td}
\end{figure*}

After explaining the bulk crystal structures, we now proceed to discussing the structures of a single layer. Here the ${monolayer-1T'}$ (${monolayer-T_d}$) is obtained by directly isolating a single layer from the bulk unit cell of the ${bulk-1T'}$ (${bulk-T_d}$) phase (Fig.~\ref{monolayer_1T_Td}). 

$\mathbf{monolayer-1T'}$: The ${monolayer-1T'}$ has two independent symmetries: a mirror plane $\mathcal{M}_{a}$ and a two-fold rotational axis $\mathcal{C}_{2a}$. The combination of these two symmetries leads to the inversion symmetry (Fig.~\ref{monolayer_1T_Td}c).

$\mathbf{monolayer-T_d}$: The ${monolayer-T_d}$ only has the mirror plane $\mathcal{M}_{a}$. Thus it can be viewed as a distortion from ${monolayer-1T'}$ where $\mathcal{C}_{2a}$ (and therefore $\mathcal{I}$) is broken. It is also interesting to note that $\mathbf{monolayer-T_d}$ lacks the $\mathcal{M}_{b}$ and $\mathcal{C}_{2c}$ symmetries that are present in $\mathbf{bulk-T_d}$ (see table~\ref{bulk_table}). This is because both $\mathcal{M}_{b}$ and $\mathcal{C}_{2c}$ are nonsymmorphic symmetries that require a translation along the $\hat{c}$ direction, which break down for a monolayer. 

In the main text, we exaggerated the the drawing of $monolayer-1T_d$ (Fig.1b) to help the readers to visualize symmetry breaking. The actual symmetry breaking is subtle which cannot be discerned by eye. Here we show the realistic atomic coordinates of the $1T_d$ phase of ${monolayer-T_d}$ WTe$_2$ in table~\ref{monolayer_table}. When $\mathcal{C}_{2a}$ is broken, the quadrilateral formed by W$^1$, W$^2$, Te$^1$ and Te$^2$ (Fig.~\ref{monolayer_1T_Td}d) is expected to deviate from a perfect parallelogram. Such a deviation can be directly seen from table~\ref{monolayer_1T_Td} because midpoints between W$^1-$W$^2$ and between Te$^1-$Te$^2$ do not overlap. By contrast, in $monolayer-1T'$ MoTe$_2$, Mo$^1$, Mo$^2$, Te$^1$ and Te$^2$ (Fig.~\ref{monolayer_1T_Td}c) form a perfect parallelogram (see table~\ref{monolayer_table}). Therefore, $\mathcal{C}_{2a}$ is a good symmetry in $monolayer-1T'$ MoTe$_2$.

\begin{center}
\begin{table}
\begin{tabular}{p{4cm}|p{5cm}|p{5cm}}
\hline
 & $monolayer-1T'$ & $monolayer-T_d$\\
\hline
Structural phase & Primitive monoclinic & Primitive monoclinic\\
Lattice constants & $a\neq{b}$ & $a\neq{b}$\\
Angles & $\alpha=90^{\circ}$ & $\alpha=90^{\circ}$\\
Space group & $P2/m(\#11)$ & $P1m1 (\#6)$\\
Point group & $C_{2h}$ & $C_{1s}$\\
Symmetries & $\mathcal{I}$, $\mathcal{C}_{2a}$, $\mathcal{M}_{a}$ & $\mathcal{M}_{a}$\\
\hline
\end{tabular}
\caption{\textbf{Key properties of the $monolayer-1T'$ and $monolayer-T_d$ structures.} $\mathcal{I}$ is the inversion symmetry; $\mathcal{C}_{2}$ is a two-fold rotational symmetry; $\mathcal{M}$ is a mirror plane.}
\label{monolayer_table}
\end{table}
\end{center}

\begin{center}
\begin{table}
\begin{tabular}{p{5cm}p{5cm}p{5cm}}
  &$monolayer-T_d$ WTe$_2$&  \\
\end{tabular}
\begin{tabular}{p{5cm}|p{5cm}|p{5cm}}
\hline
W$^{1}$ & W$^{2}$ & midpoint \\
$(0.50000, 0.96020, 0.51522)$ & $(0.00000, 0.60062, 0.50000)$ & ($0.25000, 0.78041, 0.50761$)\\
\hline
Te$^{1}$ & Te$^{2}$ & midpoint \\
$(0.00000, 0.85761, 0.65525)$ & $(0.50000, 0.70155, 0.35983)$ & ($0.25000, 0.77958, 0.50754$)\\
\hline
\end{tabular}
\begin{tabular}{p{5cm}p{5cm}p{5cm}}
  &$monolayer-1T'$ MoTe$_2$&  \\
\end{tabular}
\begin{tabular}{p{5cm}|p{5cm}|p{5cm}}
\hline
Mo$^{1}$ & Mo$^{2}$ & midpoint \\
$(0.25000, 0.68190, 0.49340)$ & $(0.75000, 0.31810, 0.50660)$ & ($0.50000, 0.50000, 0.50000$)\\
\hline
Te$^{1}$ & Te$^{2}$ & midpoint \\
$(0.75000, 0.55680, 0.35160)$ & $(0.25000, 0.44320, 0.64840)$ & ($0.50000, 0.50000, 0.50000$)\\
\hline
\end{tabular}
\caption{\textbf{Atomic coordinates of the $monolayer-T_d$ WTe$_2$ and $monolayer-1T'$ MoTe$_2$.} The data were taken from Refs.\cite{mar1992metal} and \cite{albert1992preparation}.}
\label{WTe2_monolayer_table}
\end{table}
\end{center}

\subsection*{V.2. Symmetry allowed CPGE in 2D materials} 

We provide a symmetry analysis on the CPGE in common 2D materials (graphene, MoS$_2$, WTe$_2$) mentioned in the main text in table~\ref{CPGE_2D_crystals_1}. We see that WTe$_2$ is the only one that allows CPGE with normal incidence. Graphene and MoS$_2$ systems cannot support CPGE with normal incidence even with an out-of-plane displacement field.

\begin{center}
\begin{table}
\begin{tabular}{p{5cm}|p{5cm}|p{5cm}}
\hline
 & bilayer graphene & bilayer graphene with $\vec{D}$ field\\
\hline
Structural phase & Primitive trigonal & Primitive trigonal\\
Space group & $P$-$3m1(\#164)$ & $P3m1 (\#156)$\\
Point group & $D_{3d}$ & $C_{3v}$\\
Symm. allowed CPGE & None & $\chi_{xxz}$ and $\chi_{yyz}$\\
Corresponding Exp. & None & CPGE with oblique incidence \\
\hline
\hline
 & monolayer MoS$_2$ & monolayer MoS$_2$ with $\vec{D}$ field\\
\hline
Structural phase & Primitive hexagonal & Primitive trigonal\\
Space group & $P$-$6m2 (\#187)$ & $P3m1 (\#156)$\\
Point group & $D_{3h}$ & $C_{3v}$\\
Symm. allowed CPGE & None & $\chi_{xxz}$ and $\chi_{yyz}$\\
Corresponding Exp. & None & CPGE with oblique incidence \\
\hline
\hline
 & bilayer MoS$_2$ & bilayer MoS$_2$ with $\vec{D}$ field\\
\hline
Structural phase & Primitive trigonal & Primitive trigonal\\
Space group & $P$-$3m1(\#164)$ & $P3m1 (\#156)$\\
Point group & $D_{3d}$ & $C_{3v}$\\
Symm. allowed CPGE & None & $\chi_{xxz}$ and $\chi_{yyz}$\\
Corresponding Exp. & None & CPGE with oblique incidence \\
\hline
\hline
 & monolayer $1T_d$ WTe$_2$ & monolayer $1T_d$ WTe$_2$ with $\vec{D}$ field\\
\hline
Structural phase & Primitive monoclinic & Primitive monoclinic\\
Space group & $P1m1(\#6)$ & $P1m1(\#6)$\\
Point group & $C_{1s}$ & $C_{1s}$\\
Symm. allowed CPGE & $\chi_{xxy}$, $\chi_{xxz}$, $\chi_{yyz}$ &  $\chi_{xxy}$, $\chi_{xxz}$, $\chi_{yyz}$\\
Corresponding Exp. & CPGE with normal and obilque incidence & CPGE with normal and obilque incidence \\
\hline
\end{tabular}
\caption{\textbf{Symmetry allowed CPGE in 2D materials} The bilayer graphene here refers to the usual Bernal stacking. The displacement field $\vec{D}$ is applied along the out-of-plane direction.}
\label{CPGE_2D_crystals_1}
\end{table}
\end{center}

\clearpage

\bibliography{Topological_and_2D}

\end{singlespacing}

\end{document}